\documentclass[twocolumn]{aa}    

\usepackage{epsfig,subfig}
\usepackage{amsmath}
\usepackage{natbib}
\usepackage{longtable}
\usepackage{comment}
\usepackage{color}
\usepackage{comment}

\def\spose#1{\hbox to 0pt{#1\hss}}
\def\simlt{\mathrel{\spose{\lower 3pt\hbox{$\mathchar"218$}}
     \raise 2.0pt\hbox{$\mathchar"13C$}}}
\def\simgt{\mathrel{\spose{\lower 3pt\hbox{$\mathchar"218$}}
     \raise 2.0pt\hbox{$\mathchar"13E$}}}

\usepackage{graphicx}

\usepackage{graphicx}

\begin{document}

\title{A new method to assign galaxy cluster membership using photometric redshifts}

\author{G. Castignani\inst{1,2}
\and
C. Benoist \inst{1}
}
\institute{
    Laboratoire Lagrange, Universit\'e C\^ote d'Azur, Observatoire de la C\^ote d'Azur, CNRS, 
Blvd de l'Observatoire, CS 34229, 06304 Nice cedex 4, France\\
    \and	
    Centre National d'\'{E}tudes Spatiales postdoctoral fellow, CNES, 2 Place Maurice Quentin, 75001 Paris, France\\
    \email{gcastignani@oca.eu}
}
\date{}	

\abstract
{We introduce a new effective strategy to assign {group and cluster} membership probabilities $P_{mem}$ { to galaxies using photometric redshift information. Large dynamical ranges both in halo mass and cosmic time are considered.} The method takes into account {the 
magnitude distribution of both cluster and field galaxies as well as the radial distribution of galaxies in clusters}  using a non-parametric formalism, 
and relies on Bayesian inference to take photometric redshift uncertainties into account. We successfully test the method against $1,208$ galaxy clusters within redshifts $z=0.05-2.58$ and masses $10^{13.29-14.80}~M_\odot$ drawn from wide field simulated galaxy mock catalogs mainly developed for the forthcoming {\it Euclid} mission. Median purity and completeness values of ${(55^{+17}_{-15})\%}$ and ${(95^{+5}_{-10})\%}$ are reached for galaxies brighter than 0.25$L_\ast$ within $r_{200}$ of each simulated halo and for a statistical photometric redshift accuracy $\sigma((z_s-z_p)/(1+z_s))=0.03$. The mean values $\overline{\mathsf{p}}={56}\%$ and $\overline{\mathsf{c}}={93}\%$ are consistent with the median and have negligible sub-percent uncertainties.
Accurate photometric redshifts {($\sigma((z_s-z_p)/(1+z_s))\lesssim0.05$)} and robust estimates for the cluster redshift and cluster center coordinates are required. 
The dependence of the assignments on photometric redshift accuracy, galaxy magnitude and distance from the halo center, and halo properties such as mass, richness, and redshift are investigated. Variations in the mean values of both purity and completeness are globally limited to a few percent. 
The largest departures from the mean values are found for galaxies associated with distant $z\gtrsim1.5$ halos, faint {($\sim0.25\,L_\ast$)} galaxies, and those at the outskirts 
of the halo {(at cluster-centric projected distances $\sim r_{200}$)} for which the purity is decreased, {$\Delta \mathsf{p}\simeq20\%$ at most, with respect to the mean value}.
The proposed method is applied to derive accurate richness estimates. A statistical comparison between the true ($N_{\rm true}$) vs. estimated richness ($\lambda=\sum P_{mem}$) yields on average to unbiased results, $Log(\lambda/N_{\rm true})=-0.0051\pm0.15$. { The scatter around the mean of the logarithmic difference between $\lambda$ and the halo mass is 0.10~dex for massive halos $\gtrsim10^{14.5}~M_\odot$.} 
Our estimates could therefore be useful to constrain the cluster mass function and to calibrate independent cluster mass estimates such as those obtained from weak lensing, Sunyaev-Zel’dovich, and X-ray studies. 
Our method can be applied to any list of galaxy clusters or groups in both present and forthcoming surveys such as SDSS, CFHTLS, Pan-STARRS, DES, LSST, and {\it Euclid}.}


\authorrunning{Castignani \& Benoist}
\titlerunning{A new method to assign galaxy cluster membership using photometric redshifts}

\maketitle


\section{Introduction}
Galaxy clusters and groups represent the most massive gravitationally bound structures in the Universe. High densities of both matter and galaxy counts favor the occurrence of exceptional physical phenomena such as gravitational lensing \citep{Kneib_Natarajan2011}, X-ray emission \citep{Sarazin1988,rosati2002,Bohringer_Werner2009}, and Sunyaev Zel'dovich upscattering of the cosmic microwave background due to the hot gas in the intra-cluster medium \citep{Birkinshaw1999}, spatial segregation of red and passively evolving ellipticals \citep{Poggianti2003}, star formation quenching \citep{brodwin2013}, and Active Galactic Nucleus (AGN) feedback \citep{fabian2012}. 
{ Addressing cluster membership for galaxies is crucial to understand such physical phenomena as well as for studies on galaxy evolution and cosmology. } 

{ Concerning galaxy evolution, the properties of cluster galaxies in terms of galaxy colors, morphology, and spatial segregation within the cluster core 
\citep[e.g.,][]{bassett2013,mcintosh2014}  are still debated, especially at redshifts $z\gtrsim1$ where large scale structures undergo rapid evolution and fundamental cluster galaxies features such as the tight color vs. magnitude relation known as {\it red sequence} are being established \citep[e.g.,][]{zeimann2012,santos2013,strazzullo2013,gobat2013,casasola2013,brodwin2013,zeimann2013,
alberts2013}.}

{ Concerning cosmology, cluster mass estimates are commonly inferred adopting scaling relations 
from independent X-ray \citep{Ettori2013}, Sunyaev Zel'dovich \citep{Morandi2007}, or weak lensing studies \citep{Hoekstra2013} with statistical uncertainties and systematics of a few $\sim0.1$~dex \citep[][]{giodini2013,Kohlinger2015}. } { These mass vs. observable scaling relations are then used to constrain the halo mass function 
and ultimately estimate cosmological parameters by means of differential cluster counts (per unit redshift), e.g., \citet[][]{white1993,mohr2005,rozo2010,allen2011,planckXX_2013,planckXXIV_2015,campa2015,saro2015,bocquet2016}.} { The cluster richness is also used as independent cluster mass proxy \citep[e.g.,][]{andreon2015}.
Robust membership assignments can be exploited to estimate the richness \citep{rozo2015};
however, the correct identification of both field sources and cluster members is needed. The latter is also important for robust weak-lensing mass reconstruction \citep[see][for a review]{Mellier1999}.}


Present, ongoing, and forthcoming photometric wide field surveys such as SDSS, CFHTLS, Pan-STARRS, DES, LSST, and {\it Euclid} are expected to provide increasing photometric information for distant galaxies.
Therefore strategies that apply robust membership assignments on the basis of photometric information are needed. {Nevertheless there are only a few methods that address this problem \citep[][]{brunner2000,george2011,rozo2015}.
Furthermore all of them have never been applied to samples of overdensities spanning a broad range of masses (from galaxy groups to clusters).

Moreover, to the best of our knowledge, only
\citet{brunner2000} and \citet{george2011} methods use photometric redshifts of galaxies. They both rely on specific assumptions: for example they do not consider any dependence on the distance to the cluster center when performing membership assignments.}

In the present work we introduce a new method to assign group and cluster membership to galaxies 
up to redshifts $z\sim2$ using photometric redshift information. 
The main goals of the present paper are i) introducing a new strategy and ii) testing it against a large sample of halos, extracted from a wide field galaxy mock catalog, at an unprecedented wide range of redshifts ($z\sim0-2$) and halo masses ($\sim10^{13-15}~M_\odot$).
Photometric redshifts randomized through the use of Gaussian distributions are used. The impact on the membership assignments when considering uncertainties on the cluster properties, systematics, as well as more realistic photometric redshifts will be studied in a following work.

In Section~\ref{sec:motivations_method} we outline the difficulties in assigning the membership and the motivations for a new method. In Section~\ref{sec:simulated_catalogs} we describe the simulated galaxy catalog that is used.
In Sections~\ref{sec:method} and \ref{sec:results} we introduce and apply our method to assign the membership, respectively. In Section~\ref{sec:richness_estimate} we exploit the membership probabilities to derive richness estimates.
In Section~\ref{sec:conclusions} we draw our conclusions.

Throughout this work we adopt a flat $\Lambda \rm CDM$ cosmology with matter density $\Omega_{\rm m} = 0.272$, dark energy density $\Omega_{\Lambda} = 0.728$ and Hubble constant $h=H_0/(100\, \rm km\,s^{-1}\,Mpc^{-1}) = 0.704$ { \citep{Komatsu2011}}, which are the parameters adopted in the simulations used in this work. All magnitudes are reported in the AB system \citep{Oke1974}.

Throughout the text we will refer to (simulated) clusters and halos with no distinction. However, since we are interested 
in broad redshift ($z\sim0-2$) and halo mass ($M\sim10^{13-15}M_\odot$) ranges we keep in
mind that the associated galaxy overdensities might be virialized clusters or groups, as well as still forming clusters or protoclusters.

\section{Motivation for a new method}\label{sec:motivations_method}

\subsection{Spectroscopic information}\label{sec:zspec_info}
Several studies of spectroscopically confirmed cluster and group members have been performed \citep[e.g.,][]{ramella2000,diaferio2005,biviano2013,mamon2013}. Spectroscopic confirmation of all or at least a great fraction of cluster members is nevertheless impossible since it is enormously demanding in terms of observational time and particularly challenging at high redshift ($z\gtrsim1$), even for the currently available spectrographs on 8-mt class telescopes such as VIMOS and FORS at VLT, to mention a few.
In particular, this issue greatly affects  the $z\sim1-2$ redshift range, where most of the relevant spectral features fall outside the instrumental frequency bands. For this reason the $z\sim1-2$ redshift range is commonly identified as the {\it redshift desert} \citep{steidel2004,banerji2011}.


In addition to the above mentioned problems to obtain spectroscopic redshifts for large samples of sources, especially at redshift $z\gtrsim1.5$, it is worth mentioning that spectroscopic redshifts represent in number only a small fraction ($\sim1\%$) of the photometric redshift dataset for both present and forthcoming surveys such as SDSS \citep{york2000}, DES \citep{des2005,Flaugher2005}, LSST \citep{lsst2009,lsst2012}, and {\it Euclid} \citep{laureijs2011,laureijs2014}. This is also true when surveys that have good spectroscopic coverage such as BOSS \citep{dawson2013}, GAMA \citep{baldry2014}, VIPERS \citep{garilli2014,guzzo2014}, and COSMOS \citep{scoville2008,lefevre2015} are considered.

\subsection{Photometric information}\label{sec:zphot_info}
Peculiar velocities of the galaxies result in unavoidable redshift space distortions \citep[e.g.,][]{marulli2015} and a consequent apparent elongation of clusters and groups along the line of sight $\sim0.001(1+z)$ for massive clusters of $\sim10^{14}~M_\odot$ \citep{evrard2008}. The last effect causes overmerging of distinct large scale structures as well as difficulties in disentangling field galaxies and cluster/group members along the line of sight.

These projection effects significantly affect galaxy cluster and group detections \citep{knobel2009,knobel2012,diener2013}, as well as membership assignments. This occurs also when the best spectroscopic redshift datasets available are used and peculiar velocities are carefully considered when performing the assignments \citep[for example, when the caustic method is used,][]{Diaferio1999,gifford2013,yu2015}.

Performing membership assignments on the basis of photometric information is even more challenging.
Projection effects dramatically affect the capability to separate  cluster members from foreground and background sources.
Such difficulties are ultimately due to the photometric redshift uncertainties which are significantly larger than the cluster scales, especially at the faint end of the {galaxy} luminosity function and at high ($z\gtrsim1.5$) redshifts. 
Typical statistical photometric redshift uncertainties for accurate photometric redshift estimates are in fact in the range $\sim0.03-0.05(1+z)$ for galaxies with {\sf H}-band magnitudes ${\sf H}<24$ and redshift $z\lesssim2.5$
\citep{skelton2014,ascaso2015,bezanson2016}. 

Photometric information such as colors \citep{rykoff2014,rozo2015} and/or photometric redshifts \citep{brunner2000,papovich2010,george2011} have been nevertheless widely used in previous studies to detect groups and galaxy clusters, as well as to identify their galaxy population, in particular at intermediate/high redshifts ($z\lesssim1$), where spectroscopic information is difficult to obtain for large samples of clusters and cluster galaxies.

Remarkably, recent theoretical and technical improvements in estimating redshifts using photometric information have been done. They have been achieved thanks to i) the increasing number of photometric surveys and better coverage of the electromagnetic spectrum, especially at the near infra-red \citep[e.g., UltraVISTA,][]{McCracken2012}, which is crucial for photometric redshift estimates of distant  $z>1$ sources \citep{Laigle2016}; ii) the advancements of independent techniques such as those based on Spectral Energy Distribution (SED) template fitting \citep{arnouts1999,benitez1999,bolzonella2000,ilbert2006}, machine learning techniques \citep{Collister_Lahav2004,sadeh2015,cavuoti2015}, and clustering properties \citep{menard2013,rahman2015,rahman2016}; { iii) the development of accurate (photometric vs. spectroscopic redshift) calibration strategies \citep{Cunha2012,masters2015,Newman2015}.}

In addition to all the above mentioned aspects, the advent of several on-going and forthcoming wide and/or deep multiwavelength infrared-optical-ultraviolet photometric surveys such as DES, LSST, and {\it Euclid} strongly encourages us to introduce and test against simulations a new method to perform membership assignments mainly on the basis of photometric information and in particular photometric redshifts. Before describing it in detail in the following sections we first describe the dataset used.

\section{Simulated galaxy and halo catalogs}\label{sec:simulated_catalogs}

We use the 20.4~square degree light cone galaxy mock catalog\footnote{Mock catalog { \it {Euclid\_v1\_LC\_DEEP\_Gonzalez2014a}} in the Virgo - Millennium database. Credits: {http://galaxy-catalogue.dur.ac.uk:8080/Millennium/Help?page=databases/ euclid\_v1/lc\_deep\_gonzalez2014a}}
recently developed for the {\it Euclid} consortium. 

The simulated catalog is produced using halo merger trees extracted from a N-body  $\Lambda$CDM cosmological simulation \citep{guo2013,lacey2015}.
The simulation traces $2160^3$ particles within a  cubic $500~h^{-1}$~Mpc size region from $z=127$ to the present. The halos in the simulations are populated with galaxies using the GALFORM \citep{cole2000}
semi-analytical model, { in particular, the version presented in \citet{Gonzalez-Perez2014}.}
{ This model includes those physical processes that are thought to be fundamental for understanding the formation and evolution of galaxies, such as } galactic mergers, star formation history, radiative cooling of the gas, and both supernova and AGN feedback.
{ Furthermore, an updated \citet{Bruzual_Charlot1993} stellar population synthesis model and the \citet{Kennicutt1983} initial mass function were adopted when generating the simulated catalog.}

{ The galaxy catalog used in the present work is complete down to {\it Euclid} {\sf H}-band magnitude ${\sf H}=26$ and is generated similarly to that of \citet{merson2013}, which refers to a wider and less deep survey area.} 


The final catalog thus contains useful information about galaxies such as their positions (coordinates  in the projected space and {observed} redshifts), peculiar velocities, star content and star formation rate, as well as photometric information for all galaxies in several bands of present and forthcoming surveys such as {\sf ugriz} of SDSS, {\sf grizy} of DES, and {\sf YJH} of {\it Euclid }. 

{ The {observed} redshifts included in the simulations are cosmological redshifts corrected for peculiar velocities of the galaxies. 
In this work we will use the {observed} redshifts of the mock catalog referring to them simply as spectroscopic redshifts. }

{Given the specific halo merger history included in the simulations the catalog also contains some information (such as the virial mass) about the host halo of each galaxy.
For each halo in the galaxy mock catalog we also know the exact location and properties of all its galaxy members.}


The catalog contains galaxies in the redshift range $z=0-6$ and represents a simulation of the  {\it Euclid} deep field survey which will cover approximately 40~square degrees of the sky down to {\sf Y, J, H} = 26 at 5$\sigma$ photometric accuracy. 
The combined use of multiwavelength infrared-optical-ultraviolet surveys such as SDSS, DES, and {\it Euclid} will ultimately imply photometric redshifts with an accuracy of $\sigma(\Delta z /(1+z_{s}))\lesssim0.03-0.05$
up to $z\sim2$ \citep[][and {\it Euclid} Red Book\footnote{http://sci.esa.int/euclid/48983-euclid-definition-study-report-esa-sre-2011-12/\#}]{ascaso2015}, where $\Delta z = z_{p}-z_{s}$. Here $z_{p}$ and $z_{s}$ denote photometric and spectroscopic redshifts, respectively.

\subsection{Redefinition of the galaxy catalog}\label{sec:photoz}
We consider the simulated galaxy catalog down to its completeness limit ${\sf H}=26$ and we assign simulated photometric redshifts to the galaxies, which are a fundamental ingredient of the method presented here.
The photometric redshifts are drawn from a Gaussian distribution centered at the spectroscopic redshift $z_{s}$ of each galaxy and with a standard deviation $\sigma(z_{s})=\sigma_0(1+z_{s})$. The values $\sigma_0=0.02$, 0.03, and 0.05 are chosen, so that several photometric redshift catalogs with different statistical redshift accuracy typical of real catalogs are produced. Hereafter we will consider photometric redshifts corresponding to $\sigma_0=0.03$, unless otherwise specified.
The other values will be considered for comparison.

The simplified prescription mentioned above to assign photometric redshifts is chosen in order to understand and control the impact of photometric redshift uncertainties over the wide range of redshifts and halo masses considered in this work. 
In particular, we neglect on purpose the magnitude dependence and the catastrophic failures of the photometric redshifts. The catalog  also lacks stars and quasars whose presence affects the statistical photometric redshift accuracy. We also neglect magnitude uncertainties. More realistic photometric redshifts (including bias and catastrophic failures) and datasets will be considered in future work.

We then consider only those sources that have  apparent ${\sf H}-$band magnitudes brighter than ${\sf H}_\ast(z_{p})+1.5$, i.e.,  more luminous than $\sim0.25\,L_\ast$.
Such a choice is consistent with that adopted for richness estimates, which are commonly performed using galaxies brighter than $0.4\,L_\ast$, where $L_\ast$ is the luminosity of a galaxy at the knee of the {galaxy} luminosity function  \citep{high2010,rykoff2012,jimeno2015}. 
Here ${\sf H}_\ast(z)$  is the apparent {\sf H}-band magnitude an $L_\ast$ galaxy would have if located at redshift $z$ and has been derived from the evolution of the SED of an elliptical galaxy
taken from the PEGASE2 SED library \citep{Fioc1997} and
calibrated with Coma cluster \citep{depropris1998}.
A burst of star formation at $z=5$ and an exponential decrement with time with exponent $\tau = 0.1$~Gyr are assumed. { {It was checked within the {\it Euclid} collaboration that these choices are consistent with the galaxy evolutionary model used for the simulations.}}
The final galaxy catalog comprises $1,332,513$ sources which will be used when assigning membership probabilities.

The high $\textsf{H}=26$ completeness limit assures completeness well above $z=3$ for the galaxies brighter than $\sim0.25\,L_\ast$.
Furthermore the specific redshift dependent \textsf{H}-band magnitude selection used in this work assures that we are not rejecting bright high-$z$ early type sources, which would be instead rejected in the case of a different selection, for example in \textsf{i}-band. For the last case  the 4,000~\AA~ break in the rest frame SED of ellipticals at $z\gtrsim1$ is in fact redshifted at wavelengths that are longer than the characteristic wavelength of the \textsf{i}-band filter.

The adopted ${\sf H}_\ast(z_{p})+1.5$ magnitude cut is also motivated by the two following observational facts:  
i) photometric redshift accuracy has a strong dependence on magnitude. In fact photometric redshifts undergo catastrophic failures at faint magnitudes \citep{george2011,bezanson2016}. ii) Bright, red, and elliptical sources are expected to populate the central regions of groups and clusters and to occupy a tight region in color-magnitude plots known as red sequence. Although the presence and evolution of this segregation and of the red-sequence is still debated and not fully understood, in particular at $z\gtrsim1.4$, sources with luminosity around the characteristic $L_\ast$ luminosity of the {galaxy} luminosity function 
represent a consistent fraction of the cluster galaxy population.

The main goal of this project is the future application of our method to real datasets and therefore, because of all the above mentioned aspects, we prefer to maintain the magnitude cut described above even if, limited to this work, we are considering simplified photometric redshift assignments.


\subsection{Redefinition of the halo catalogs}\label{sec:halo_cat_redef} 
An important step for our strategy is the selection of the cluster center, which is a key ingredient of many studies on galaxy clusters \citep[e.g.,][and references therein]{cui2016,rossetti2016}.
For each halo we consider its members { which have magnitudes brighter than} ${\sf H}_\ast(z_{s})+1.5$ { and we estimate its barycenter averaging the Cartesian coordinates of these members.} 
We use the halo center coordinates to estimate i) the halo redshift as the cosmological redshift associated with the barycenter and ii) the halo center coordinates as the ra-dec coordinates of the barycenter.
{ {We note that the simulated galaxy catalog contains information about the central galaxy of the halo. However we preferred to re-estimate the cluster center as the barycenter of the cluster members because this is closer to the position of the number density peak of galaxies which cluster finders tend to detect.}}

Then we use the halo redshift and its virial mass to estimate the corresponding virial radius (denoted hereafter as $r_{200}$)
as the radius at which the enclosed virial mass encompasses the matter density 200 times the critical one at the halo redshift.

{ We stress that the mock catalog contains the exact positions of all galaxies as well as cluster membership information. Cluster members, {i.e., all galaxies that are located in a simulated halo},  are in fact directly identified when generating the simulations following the merging history of the halos. {We refer to \citet{merson2013} for more detail.}}

{ The virial mass included in the simulations does not exactly correspond to the halo mass $M_{200}$ \citep{Jiang2014}. 
Nevertheless, we will always refer to halo masses and radii as $M_{200}$ and $r_{200}$, respectively.
This approximation results in typical uncertainties of $\lesssim25\%$ and $\lesssim8\%$ with respect to the correct values, respectively \citep{Jiang2014}.}

{In order to test the membership method presented in this work we restrict to those clusters which are safely within the survey area}, i.e., those halos whose centers lie at least 5~Mpc from the light cone boundaries. The 5~Mpc radius is chosen in order to perform local galaxy number density estimates (see Section~\ref{sec:number_densities}). We also restrict our analysis to halos with masses $\geq10^{13}~M_\odot$, which are typical of groups and clusters and represent the masses the halo mass function is most sensitive to \citep{bode2001}.

In order to ensure good statistics and robust membership assignments we further limit ourselves to those halos which have at least 10 members in the final photometric redshift galaxy catalog { with a projected distance from the cluster center not greater than $r_{200}$}. 
Our final sample comprises $1,208$ halos within the range of redshifts $z=0.05-2.58$ and halo masses $Log(M/M_\odot)=13.29-14.80$.
The median logarithmic mass is $Log(M/M_\odot)= 13.87\pm0.24$ and the median redshift is $z=1.00\pm0.47$, where the reported uncertainties denote the root mean square (rms) dispersion.

\begin{figure*} \centering
\subfloat{\includegraphics[width=0.5\textwidth]{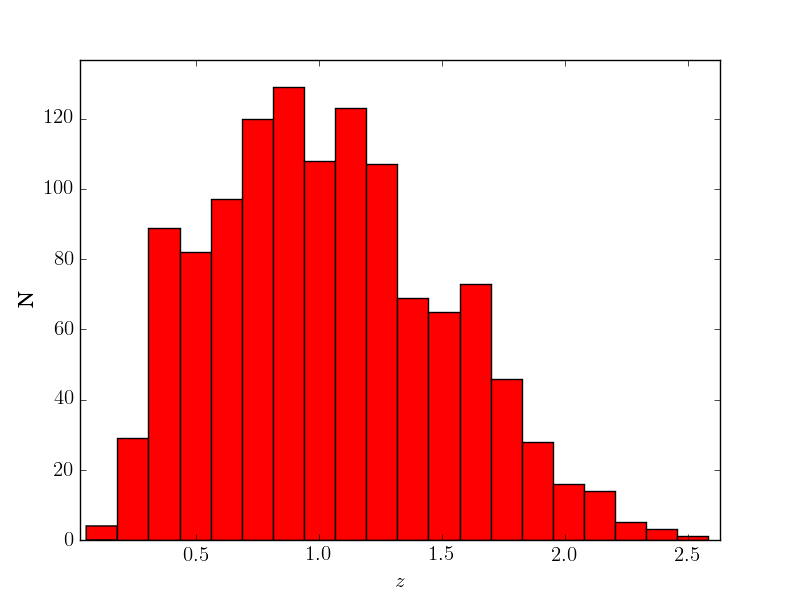}}\qquad
\subfloat{\includegraphics[width=0.5\textwidth]{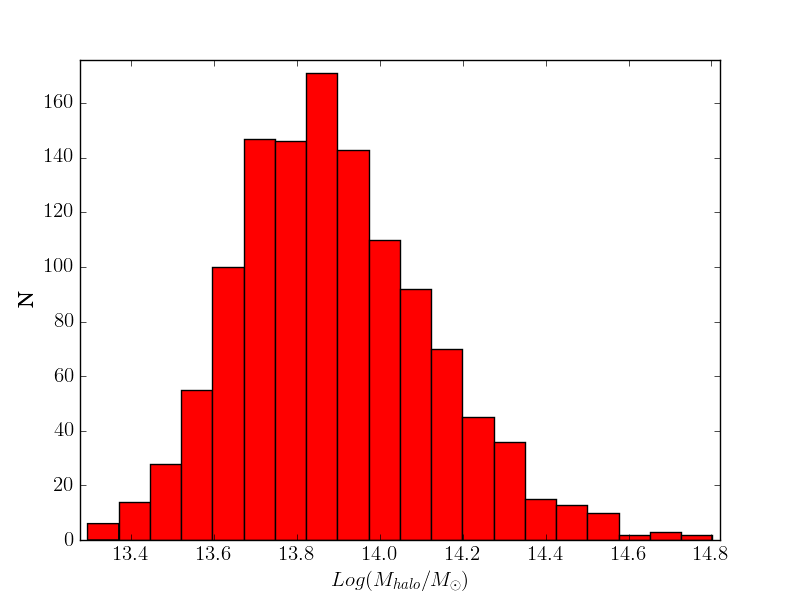}}\qquad
\subfloat{\includegraphics[width=0.5\textwidth]{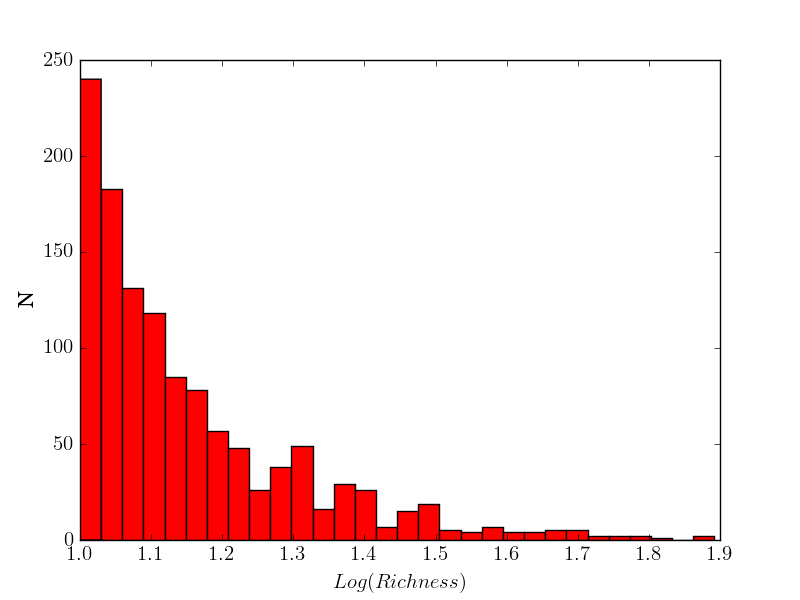}}\qquad
\caption{From top to bottom. Redshift, mass, and richness distributions for the halos in the sample. { Richness refers to cluster galaxies brighter than ${\sf H}_\ast(z_{p})+1.5$ and within $r_{200}$ radius of each halo.}}
\label{fig:halo_histo}
\end{figure*}

In Figure~\ref{fig:halo_histo} we show the distribution in redshift, mass, and richness of the halos in the sample. { Richness refers to cluster galaxies brighter than ${\sf H}_\ast(z_{p})+1.5$ and 
with a projected distance from the cluster center not greater than $r_{200}$.}
Because of the limited area of the survey the number statistics is poor at high and low redshifts, as well as at high and low halo masses. Concerning the last ones, the statistics is suppressed by the specific richness cut applied.

Concerning redshifts $z\gtrsim2$ we also consider our results cautiously. This is mainly because the properties of galaxy clusters and groups are highly uncertain at these redshifts \citep{toshikawa2014,kubo2015,diener2015,muldrew2015}, where forming galaxy (proto)clusters are expected to be associated with a significant fraction of the most massive halos. 

\section{Method}\label{sec:method}
In this Section we introduce a new method to perform robust membership assignments on the basis mainly of photometric redshift information.

We have been inspired by the work of \citet{rykoff2012} and by their subsequent studies \citep[][]{rykoff2014,rozo2015} as well as by \citet{george2011}.
We adopt a similar probabilistic Bayesian formalism to assign membership probabilities. However, through the use of Bayesian inference, we attempt an effective generalization of their work. We use in our method the available information about halos and galaxies within the cluster virial radius: cluster redshifts and cluster center coordinates, as well as coordinates, magnitudes, and photometric redshifts of galaxies.

As discussed below we adopt an operative (almost non-parametric) approach based on galaxy number counts which are estimated locally in the cluster field at given  redshift and magnitude bins. This strategy is preferred to that which assumes the {\it a priori} knowledge of specific models such as those describing the galaxy luminosity function and cluster radial profiles.
Our choice is mainly motivated by our ignorance of the cluster galaxy luminosity function and the cluster radial profiles over the broad range of halo masses considered, especially at redshifts $z\gtrsim1$, where strong evolution of both  megaparsec-scale structures and cluster galaxies occurs.
{ Furthermore, assuming a specific cluster profile could lead to a bias if clusters with relaxed and disturbed morphology are not considered separately.}

\subsection{Catalogs of galaxies and halos}\label{sec:catalog_priors}
We stress that the method proposed in this paper is not a method to detect clusters of galaxies but a method {to assign robust group and cluster membership to galaxies.}
In full generality we consider a magnitude limited catalog of galaxies (each of them is denoted with the letter $g$) and a catalog of detected (or simulated) groups and/or clusters (each of them is denoted with the letter $c$), similarly to those used in this work.

For each galaxy we consider i) the observed magnitude $m_g$ in a given reference band, 
ii) the right ascension - declination coordinates (ra$_g$,dec$_g$), and iii) the photometric redshift. 


Similarly, for each group/cluster we consider i) the right ascension - declination coordinates (ra$_c$,dec$_c$) of the cluster center (i.e., the barycenter, which is estimated from the cluster members in our case), ii) the radius $r_{200}$
of the cluster or, alternatively, an estimate of the cluster size in physical units, and iii) the 
probability density function (PDF) associated with the cluster redshift, $P_c(z)$. All three quantities can be estimated and/or are provided in group/cluster catalogs.

\subsection{Adopted strategy}\label{sec:adopted_strategy}
The method we introduce is mainly based on photometric redshifts and galaxy number counts. Similarly to other methods that use photometric redshift information \citep[e.g.,][and references therein]{Eisenhardt2008,Bellagamba2011,Castignani2014a,Castignani2014b}
to search for or study galaxy clusters and groups, we consider the redshift information and the coordinates in the projected space separately.  

Core sizes are typically in the range 0.1-0.4~Mpc for rich clusters \citep{Bahcall1975,Dressler1978,Sarazin1986}, while photometric redshift uncertainties of $\pm0.03(1+z)$ correspond to $\sim1.0$, 0.7, and 0.5~$\times10^2$~Mpc at redshifts $z=0.5$, $1.0$, and $2.$, respectively.

Therefore photometric redshift uncertainties are much larger (by a factor of $\sim100$) than the typical scale of the cores of clusters and groups and are in fact significantly dominant with respect to any other observable uncertainty (e.g., flux uncertainties, projected space coordinate uncertainties). 

A detailed distance discrimination based on photometric redshifts
is therefore needed. As it will be clarified below this can be
achieved to the detriment of a less detailed tessellation of the
projected space.

We adopt a treatment of the projected space (i.e., counts in cells/shells) and the photometric redshift information (i.e., counts in redshift bins) similar to that used in the Poisson Probability Method \citep[PPM,][]{Castignani2014a,Castignani2014b}
which was introduced and applied to search for distant galaxy clusters and groups around a specific point in the sky using photometric redshifts of galaxies and galaxy number counts.

Before introducing our method in the following we will focus on the redshift information.

\subsubsection{Redshift information}\label{sec:redshift_information}
First we carefully consider, for each galaxy $g$, the PDF $P_g(z)$ which tells the probability that the spectroscopic redshift of the galaxy $z_{s,g}$ is in the range  $(z;z+\delta z)$. 
{In the case of galaxies for which the spectroscopic redshift is known, $P_g(z)$ is reduced to a very narrow distribution centered at the spectroscopic redshift.}

Since we are always considering photometric redshifts, $z_{p,g}$, we rewrite in a compact form the PDF as the conditional probability distribution $P(z_{s,g}|z_{p,g})$.
The Bayes theorem allows us to relate the PDF to the galaxy redshift distribution $N(z_{s})=dN/dz_{s}$ and the PDF $P(z_{p,g}|z_{s,g})$, which is, by construction, equal to a Gaussian $\mathcal{N}(z_{p,g},\mu,\sigma)$ with mean $\mu=z_{s,g}$ and sigma $\sigma=\sigma_0(1+z_{s,g})$. It holds:
\small
\begin{equation}
  \label{eq:prob1}
 P(z_{s,g}|z_{p,g})\propto N(z_{s})\cdot\mathcal{N}(z_{p,g},\mu=z_{s,g},\sigma=\sigma_0(1+z_{s,g}))\;,
\end{equation}
\normalsize
where the normalization is fixed requiring that the integral in $z_{s,g}$ is one.
We refer to \citet{Sheth_Rossi2010}, where the same equation is derived (their Equation~1).

An interesting consequence of Equation~(\ref{eq:prob1}) is that, even if photometric redshifts are assigned with a prescription which is independent of the magnitude { (by construction, $\mathcal{N}(z_{p,g},\mu,\sigma)$ in our case of Gaussian photometric redshifts does not depend on magnitude)}, the actual functional form of  $P(z_{s,g}|z_{p,g})$ is magnitude dependent because of the presence of the redshift distribution $N(z_{s})$, which is function of the specific magnitude of the galaxy considered.

Furthermore $P(z_{s,g}|z_{p,g})$ also depends on the local clustering properties.
This is because we are particularly interested in those galaxies that are {within the cluster virial radius.} The dependence on clustering properties is implicitly present in $N(z_{s})$, which is in fact the local redshift distribution in the cluster field.

Several previous studies developed similar formalism to study in detail the statistical properties of galaxies in redshift surveys \citep{Efstathiou1988,Sheth2007,Benjamin2007,Fu2008}.
However in our case we are interested in estimating locally the redshift distribution $N(z_{s})$.
Therefore we prefer not to estimate the redshift distribution using the entire survey, which would lead to a possible underestimation of the number counts in the case of galaxies {within the cluster virial radius.}
On the other hand if we estimated the redshift distribution using local number counts (e.g., counts in cells) we would be highly affected by low number counts and shot noise, especially at the bright end of the {galaxy} luminosity function.

Motivated by these aspects, in order not to introduce any artificial bias and systematics in the estimate of $P(z_{s,g}|z_{p,g})$ in terms of its peak and shape we prefer to consider conservatively a constant redshift distribution $N(z_{s})$. As it is clear from Equation~(\ref{eq:prob1}) such a choice relies implicitly on the assumption that $N(z_{s})$ does not vary dramatically for $z_{s}\approx z_{p,g}$ or, more precisely, over the support of $P(z_{s,g}|z_{p,g})$.  

Given the arguments outlined above we therefore derive our final expression for $P_g(z)$ as:
\begin{equation}
 \label{eq:prob2}
 P_g(z) \propto\frac{1}{\sigma_0(1+z)}exp \Bigl[ -\frac{(z-z_{p,g})^2}{2\sigma_0^2(1+z)^2}\Bigr]\;,
\end{equation}
where the normalization is again fixed requiring that the integration in $z$ is equal to one. 

\begin{figure} \centering
\includegraphics[width=0.5\textwidth]{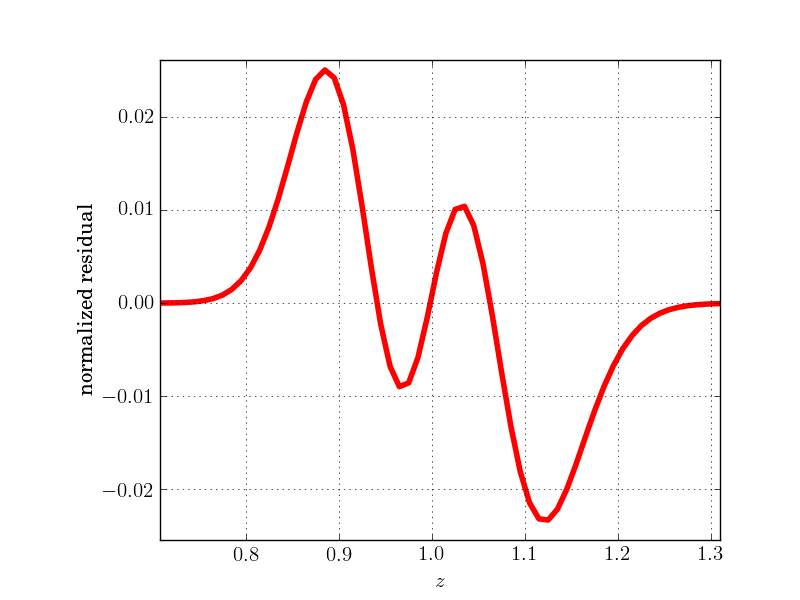}\qquad
\caption{Graphical comparison between $P_g(z)$ as in Equation~(\ref{eq:prob2}) and its Gaussian approximation $\mathcal{N}(z,\mu=z_{p,g},\sigma=\sigma_0(1+z_{p,g}))$, where $z_{p,g}=1$ is chosen equal to the median redshift of the halos in the sample.
Normalized residuals $(\mathcal{N}(z)-P_g(z))/{\sf max}(\mathcal{N}(z))$ are plotted as a function of $z$. By construction the integral of the function is null.}
\label{fig:PDF_residuals}
\end{figure}

We stress that, while $P(z_{p,g}|z_{s,g})$ is, by construction, symmetric around $z_{s,g}$, $P_g(z_{s,g})\equiv P(z_{s,g}|z_{p,g})$ is not symmetric around $z_{p,g}$.
In Figure~\ref{fig:PDF_residuals} we show a graphical comparison between $P(z_{s,g}|z_{p,g})$ and its  Gaussian approximation $\mathcal{N}(z_{s,g},\mu=z_{p,g},\sigma=\sigma_0(1+z_{p,g}))$.
When compared to the Gaussian approximation the PDF reported in Equation~(\ref{eq:prob2}) shows a shift in redshift of the peak towards lower redshifts and an excess at high redshifts, as also clear from Equation~(\ref{eq:prob2}). 

The distortion with respect to the Gaussian approximation is in fact due to the $\propto\sigma_0(1+z)$ scaling of the statistical redshift uncertainties, which are higher at increasing redshifts and imply an increasing spread of the photometric redshift distribution with increasing spectroscopic redshifts.

While the correction is relatively small (i.e., at percent level), we stress that it plays a role in determining membership assignments more accurately, which is particularly important in the context of (future) high-precision cosmological studies.
Our findings are also consistent with previous work by \citet{Sheth_Rossi2010}, who found discrepancies when comparing statistically the two PDFs, $P(z_{s,g}|z_{p,g})$ and $P(z_{p,g}|z_{s,g})$, when drawn from the SDSS survey.

In order to apply our formalism to the case of membership assignments, we consider the following prescriptions, concerning the redshift space, as in \citet{george2011}.
For practical reasons we discretize $P_g(z)$ given in Equation~(\ref{eq:prob2}) within the redshift range $z=0-3$, which safely includes all redshifts of the halos considered in our sample. We also use consecutive and finite redshift bins $\delta z=0.01$, which assure a good redshift sampling in the case of typical photometric redshift surveys.
In order to avoid a spiky behavior of $P_g(z)$ due to the adopted discrete binning we also safely convolve $P_g(z)$ with a Gaussian $\mathcal{N}(z,\mu=0,\sigma=\delta z)$ which implies a suppression of 
high-frequency $\gtrsim1/\delta z$ fluctuations.

In full analogy with the formalism outlined, for each halo $c$, we assume  $P_c(z)$ to be a Gaussian, $\mathcal{N}(z,\mu=z_c,\sigma=\sigma_c(1+z_c))$, where $z_c$ denotes here some estimate for the cluster redshift. { Galaxy clusters are detected from photometric redshift surveys with a typical {statistical} redshift accuracy $\sigma_c\simeq\sigma_0/2$ \citep{Wen2009,Wen2012}. We nevertheless assume $\sigma_c=\sigma_0$. This represents a conservative choice.
{Concerning our case of simulated Gaussian photometric redshifts our choice} ultimately favors slightly higher values of completeness despite of slightly lower values of purity for our membership assignments. 
We refer to Section~\ref{sec:cp_plot} and Figure~\ref{fig:cp_multi_panels} (bottom right panel), where  the impact of choosing different values of $\sigma_c$ is tested.}

Then we discretize $P_c(z)$ and remove high-frequency $\gtrsim1/\delta z$ fluctuations applying a Gaussian convolution, analogously to what has been done for $P_g(z)$. 

We will always consider $z_c$ equal to the value of the halo redshift, estimated in Section~\ref{sec:halo_cat_redef}. Even if uncertainties could affect the estimates of the cluster redshift as well as of all other cluster properties we point out that we prefer not to include them in this work. This is mainly because one of the main goals of this work is to introduce and test our method against photometric redshift catalogs under controlled statistical uncertainties which are limited to the galaxy catalog. The impact of systematics and uncertainties on the cluster properties will be studied in a following work.

We also stress that the above outlined strategy is tailored to the specific properties of the photometric catalog considered and in particular to the prescription used to assign photometric redshifts. A different strategy could be possibly applied in the case of more realistic photometric redshift catalogs which include more complex statistical as well as systematic uncertainties.


\subsubsection{Photometry}
Accordingly to the galaxy catalog redefinition performed in Section~\ref{sec:photoz}, in order to perform membership assignments we consider the photometry of the galaxies in $\mathsf{H}$-band, which is the reference band of our catalog. 
We prefer not to use additional information in other bands such as color information and bivariate {galaxy} luminosity functions. 
In fact, since our main goal is to perform membership assignments for galaxies in groups and clusters over a broad range of redshifts and cluster masses we do not want to be biased towards specific galaxy colors, whose distribution and evolution with both redshift and cluster mass are still debated, especially at $z\gtrsim1$. 

In full analogy with the strategy adopted for the redshift information, for each galaxy $g$ we define the PDF $P'_g(m)$ which tells the probability that the \textsf{H}-band magnitude of the galaxy is in the range $(m;m+\delta m)$. We also discretize the problem considering consecutive and discrete bins $\delta m=0.1$ down to $\mathsf{H}=26$ and conservatively assume $P'_g(m)=\mathcal{N}(m,\mu=m_g,\sigma=\delta m)$, where $m_g$ is the observed magnitude of the galaxy. The adopted $\sigma$ if equal to the bin size, which is on the order of the typical statistical photometric uncertainties we expect. Therefore our formalism effectively reproduces the statistical magnitude uncertainties and suppresses high-frequency $\gtrsim1/\delta m$ noise in the discrete PDF.

\subsubsection{Number densities}\label{sec:number_densities}
Within the framework described in the previous Sections we introduce here the mean background number density $N_{bkg}(m,z)$ which is the mean galaxy number counts per unit redshift, magnitude, and solid angle. 
The background densities are estimated both globally (i.e., considering a $\Omega=10.61$~square degree rectangular area inscribed in the light cone of the survey) and locally (i.e., considering for each halo the annulus comprised within 3 and 5~Mpc from the halo center). We refer to the two estimates as  $N_{bkg}^{glob}$ and $N_{bkg}^{loc}$, respectively.

We can define $N_{bkg}^{glob}(m,z)=\frac{1}{\Omega}\cdot\sum_gP_g(z)P_g(m)$, where the sum is performed over all galaxies which are within the rectangular area considered. 
$N_{bkg}^{loc}$ is defined analogously, where the galaxies in each annulus and the associated area
are considered. { The (discrete) PDFs $P_g(z)$ and $P_g(m)$ satisfy the condition $\sum_i P_g(z_i) = \sum_jP_g(m_j)=1$, where the summation is performed over bins of redshift ($\delta z$) and magnitude ($\delta m$) centered at $z_i$ and $m_j$, respectively. }

Remarkably, the delocalization in redshift through the use of the PDF partially 
overcomes some problems originated by photometric redshift uncertainties such as the difficulty and ambiguity in determining physical (redshift dependent) distances among sources.
In fact the PDF in redshift provides us a general tool to define number counts, as well as to convert the subtended solid angles into physical areas and, therefore, number counts into number densities.

{ Then we define the following running mean integrating $N_{bkg}^{glob}$ both in redshift and magnitude. This is done to avoid very low number counts which are originated by the small magnitude and redshift bins adopted. It holds:}
\small
\begin{equation}
\nonumber
\langle N_{bkg}^{glob}(m,z)\rangle = \int_{m-5\delta m}^{m+5\delta m}\int_{z-2\sigma(z)}^{z+2\sigma(z)}N_{bkg}^{glob}(m',z')\;dz'dm'\;,
\end{equation}
\normalsize
where $\sigma(z)=\sigma_0(1+z)$ denotes the 1-$\sigma$ statistical redshift uncertainty.
An analogous definition could be introduced for $N_{bkg}^{loc}$. However, because of the limited area adopted for the local background selection we may be still affected by small number counts, especially at the bright end of the luminosity function. Therefore, when estimating $\langle N_{bkg}^{loc}(m,z)\rangle$ we  prefer to adopt the same functional form as in $\langle N_{bkg}^{glob}(m,z)\rangle$, normalizing for the number counts down to $\mathsf{H}_\ast(z)+1.5$, as follows: 
\begin{equation}
\label{eq:Nloc}
\langle N_{bkg}^{loc}(m,z)\rangle = f\cdot\langle N_{bkg}^{glob}(m,z)\rangle\;,
\end{equation}
where
\small
\begin{equation}
 \label{eq:f-factor}
f = \frac{\int_{m'\leq\mathsf{H}_\ast(z)+1.5}\int_{z-2\sigma(z)}^{z+2\sigma(z)}N_{bkg}^{loc}(m',z')\;dz'dm'}{\int_{m'\leq\mathsf{H}_\ast(z)+1.5}\int_{z-2\sigma(z)}^{z+2\sigma(z)}N_{bkg}^{glob}(m',z')\;dz'dm'}\;.
\end{equation}
\normalsize

The running means $\langle N_{bkg}^{loc}(m,z)\rangle$ and $\langle N_{bkg}^{glob}(m,z)\rangle$ are crucial quantities for our membership assignments.

\begin{figure} \centering
\includegraphics[width=0.5\textwidth]{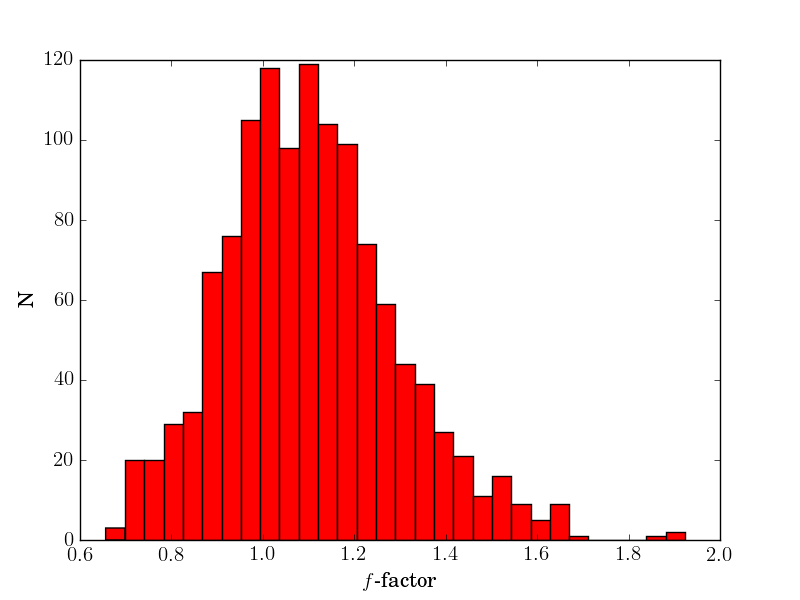}\qquad
\caption{Distribution of the $f$-factor { for the halos in the sample},  estimated at the cluster redshift.}
\label{fig:f_factorhisto}
\end{figure}

In Figure~\ref{fig:f_factorhisto} we plot the distribution of the $f$-factor { for the halos in the sample.}
The median of the distribution is in $1.09\pm0.19$, consistent with $f=1$ within the reported (relatively small) rms dispersion. 
The fact that the distribution is centered at a value slightly higher than one (the mean value is 1.10) { might be explained by the fact that the local background probes scales that are smaller than those of the global background and are therefore characterized by higher clustering.} 

Even if $\langle N_{bkg}^{loc}(m,z)\rangle\simeq\langle N_{bkg}^{glob}(m,z)\rangle$ we nevertheless prefer to use $\langle N_{bkg}^{loc}(m,z)\rangle$ (unless otherwise specified) because it better traces the local density around the cluster. It is in fact commonly used when estimating the cluster size from real surveys.
Furthermore, in the case of pointed observations of clusters it is the only one which is available.

For each halo $c$ we also introduce the local number density $N_{tot,c}(m,z,r)$ which is equal to the galaxy number counts per unit redshift, magnitude, and solid angle at a given projected physical separation within $(r;r+dr)$ from the halo center. Here, for each halo we assume azimuthal symmetry around the axis connecting the observer and the the halo center. This choice is mainly due to the need of sufficient number statistics that is obtained through the use of counts in cells/shells. 
Our assumption is also motivated by previous statistical studies, which widely used azimuthal symmetry in determining cluster radial profiles and luminosity functions of cluster galaxies \citep{biviano2013}. 
Similarly to what has been done for the background, $N_{tot,c}(m,z,r)$ is derived using the full PDFs in redshifts and magnitudes, while positional uncertainties are neglected. In analogy with the background, in order to limit shot noise fluctuations, we similarly define a running mean as follows:
\footnotesize
\begin{align}
\label{eq:Ntot}
\langle N_{tot,c}(m,z,r)\rangle = \int_{m-5\delta m}^{m+5\delta m}\int_{z-2\sigma(z)}^{z+2\sigma(z)}\int_{r_<}^{r_>} N_{tot,c}(m',z',r')\\
\;\nonumber  dr'dz'dm'\;,
\end{align}
\normalsize
where $r_<$ and $r_>$ are the projected separations which define a subtended area, centered at $r$, equivalent to that of a circle of 450~kpc radius at the redshift $z$. This size is typical of the core of rich groups and clusters.

We stress again here that, because of the delocalization in redshift of each galaxy through the use of the full PDF, both  $N_{tot,c}(m,z,r)$ and $N_{bkg}(m,z)$ are well defined quantities.

Using the formalism developed in the present and previous Sections in the following we describe our procedure to assign membership probabilities.

\subsection{Membership probability}\label{sec:Pmem}
We  consider a specific cluster/group $c$ and a galaxy $g$ in its field and we ask which is the probability $\mathcal{P}$ that the galaxy $g$ belongs to the cluster $c$, i.e., $g\in c$.

All information outlined in the previous Sections, useful when assigning membership, is summarized in a more compact form as:
\begin{align}
\Pi = \{r_{c,g}\,;\; P_g(z)\,;\; P'_g(m)\,;\; {\rm ra}_c\,;\; {\rm dec}_c\,;\; P_c(z)\,;\;\nonumber \\
N_{tot,c}(m,z,r_{c,g}),N_{bkg,c}^{loc}(m,z)\}\;, 
\end{align}
where $r_{c,g}(z)$ is the projected physical distance between the cluster center and the galaxy coordinates. 
The pedices $g$ and $c$ simply denote that we are referring to the galaxy $g$ and the halo $c$, respectively.

The spirit behind this work is to assign membership using a distance discrimination based on photometric redshifts (see also Section~\ref{sec:adopted_strategy}). Therefore, we use Bayesian inference and express the membership probability putting emphasis on the redshift information as follows:

\small
\begin{equation}
\label{eq:pmem1}
\mathcal{P}(g\in c|\Pi) = \int \mathcal{P}(g\in c| z'_c,z'_{s,g},m'_g,\Pi)d\mathcal{P}(z'_c,z'_{s,g},m'_g|\Pi)\;,
\end{equation}
\normalsize

where $\mathcal{P}(A|B)$ is the conditional probability of the event A given B.
The posterior probability distribution in the integrand is simply\footnote{For the sake of clarity
the prior $\Pi$ is omitted in the right-hand side of the expression.} $d\mathcal{P}(z'_c,z'_{s,g},m'_g|\Pi)=P'_g(m'_g)P_g(z'_{s,g}|z'_{c})P_c(z'_{c})\,dz'_c\,dz'_{s,g}\,dm'_g$, where we consider independently the magnitude information of the galaxies, as well as the redshift information for both the galaxy and the cluster, i.e.,  $P_g(z'_{s,g}|z'_{c}) = P_g(z'_{s,g})$.

We estimate $\mathcal{P}(g\in c| z'_c,z'_{s,g},m'_g,\Pi)$, i.e., the probability that the galaxy belongs to the cluster knowing the redshift $z'_c$ of the cluster, the spectroscopic redshift ($z'_{s,g}$), and the magnitude ($m'_g$) of the galaxy, as well as all information in $\Pi$. Such a probability is given by the following number count excess:

\small
\begin{equation}
\label{eq:pmem2}
\mathcal{P}(g\in c| z'_c,z'_{s,g},\Pi) = \bigg[ 1- \frac{N_{bkg,c}^{loc}(m'_g,z'_c)}{N_{tot,c}(m'_g,z'_c,r_{c,g})}\bigg]\phi(z'_{c},z'_{s,g})\;,
\end{equation}
\normalsize

where $\phi(z'_c,z'_{s,g})$ is a general non-negative function which is positive { and less or equal to one} for $|z'_c-z'_{s,g}| \lesssim \delta z(c,g)$. Here $\delta z(c,g)$ is a generic function of cluster and/or galaxy properties. Its value is determined by the velocity dispersion of galaxies in the cluster,
i.e., $\lesssim2,000$~km~s$^{-1}$, equivalently $\delta z(c,g)\lesssim0.007(1+z_c)$ \citep{evrard2008}. As mentioned in Section~\ref{sec:zphot_info} such a dispersion is however much smaller than typical photometric redshift uncertainties. The function $\phi(z'_c,z'_{s,g})$ thus approximately reduces in our case to a { delta Kronecker} $\delta_{z'_c,z'_{s,g}}$, { which is equal to one whenever both $z'_c$ and $z'_{s,g}$  belong to the same redshift bin of width $\delta z$.}
Under the framework outlined above we provide the following expression for the membership probability:
\tiny
\begin{equation}
\label{eq:pmem5}
\mathcal{P}(g\in c|\Pi) = \int\bigg[ 1- \frac{N_{bkg,c}^{loc}(m'_g,z'_c)}{N_{tot,c}(m'_g,z'_c,r_{c,g})}\bigg]\phi(z'_{c},z'_{s,g})d\mathcal{P}(z'_c,z'_{s,g},m'_g|\Pi)\\
\end{equation}
\normalsize

The equation suggests that the membership probability can be also interpreted as the averaged number count excess, where the average is performed with the posterior probability distribution as probability measure.  

As mentioned in Section~\ref{sec:number_densities} small number counts affect number densities which motivate us to factorize the number count excess out of the integral and approximate the membership probability as follows:
\begin{align}
\label{eq:pmem5tris}
\mathcal{P}(g\in c|\Pi)\simeq (1- \beta)\int \phi(z_c',z_{s,g}')P_g(z_{s,g}')P_c(z_c')\,dz_{s,g}'~dz_c'\nonumber\\
\simeq (1- \beta)\delta z 
\int P_g(z)P_c(z)\,dz   \nonumber\\
\simeq  (1- \beta)\sum_i P_g(z_i)P_c(z_i)\,,
\end{align}
{ where the last expression explicitly shows the discretization in redshift adopted in the formalism. The Kronecker delta approximation of $\phi$ is also exploited.
The term: } 
\begin{align}
\label{eq:pmem5bis}
\beta = \frac{\langle N_{bkg,c}^{loc}(m_g,z_c)\rangle}{\langle N_{tot,c}(m_g,z_c,r_{c,g})\rangle}
\end{align}
is such that $1-\beta$ is the number count excess that is factorized out of the integral of Equation~(\ref{eq:pmem5}) and is assumed equal to a constant value.
Such a value is estimated through the use of average number counts evaluated at the cluster redshift $z_c$ and the magnitude $m_g$ of the galaxy. The local number density $N_{tot,c}$ is also evaluated at the projected separation $r_{c,g}$ of the galaxy from the cluster center, as in Equation~(\ref{eq:Ntot}).

Furthermore, number counts are averaged in radius ($r_<\leq r\leq r_>$), magnitude ($|m'_g-m_g|<5\delta m$), and redshift ($|z'_c-z_c|<2\sigma_0(1+z_c)$), consistently with the above definitions for $\langle N_{bkg,c}^{loc}\rangle$ and $\langle N_{tot,c}\rangle$.
Such a filtering  allows us also to mitigate possible statistical uncertainties such as those on cluster center coordinates, on photometry, and on photometric redshift over scales smaller than those associated with the average procedure.

The probability is set to zero in those cases where $\langle N_{bkg,c}^{loc}(m_g,z_c)\rangle \geq \langle N_{tot,c}(m_g,z_c,r_{c,g})\rangle$.
Excluding these cases and limiting ourselves to true halo members the median and mean number count excess 1-$\beta$, and the rms dispersion around the median are 0.72, 0.68, and 0.19, respectively.

Through the use of Equation~(\ref{eq:pmem5}) we operatively { take into account} 
in our formalism i) the { magnitude distribution} of field galaxies, ii) the { magnitude distribution} of cluster galaxies, and iii) the radial distribution of galaxies in clusters. 
{ The three points are exploited through the use of number counts at different bins in magnitude and distance to the cluster center and without assuming specific models for the luminosity function and the cluster profile. 
Nevertheless, membership probabilities could be used to estimate {\it  a posteriori}
both the luminosity function {of cluster galaxies} and the cluster profile 
\citep[see e.g.,][]{Dahlen2002,Dahlen2004}.}


Although the main goal of this paper is to introduce our method and test it with simulations we point out that Equation~(\ref{eq:pmem5}) is a fully general formula for the membership probability that can be applied to any list of clusters and photometric redshift catalog of galaxies.
Any systematic or statistical uncertainty as well as additional information such as redshift bias, spectroscopic redshifts, and positional uncertainties can be incorporated in our Bayesian formalism through the use of additional posterior distributions.

\subsubsection{From relative to absolute probabilities}\label{sec:Pmem_rescaling}
\begin{figure}[htbp]
\includegraphics[width=0.5\textwidth]{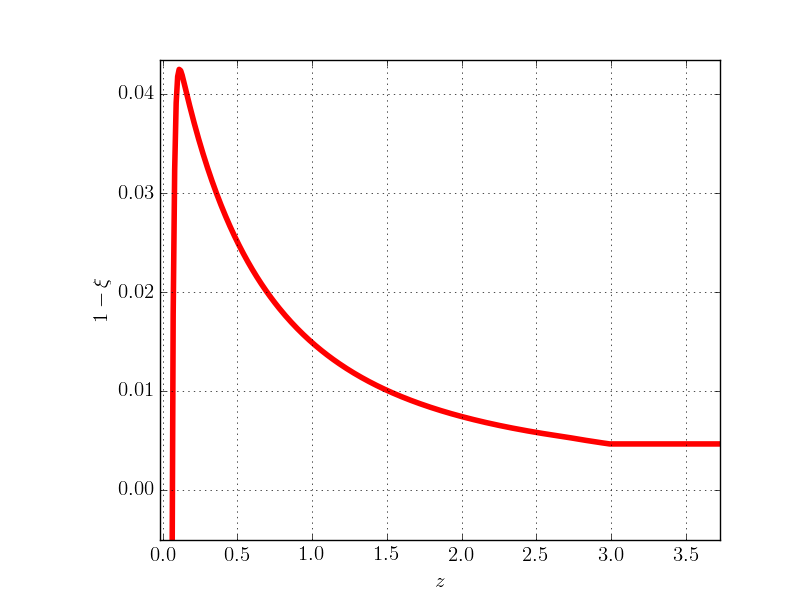}
\caption{\small Correction $1-\xi$ as a function of redshift, see Equation~(\ref{eq:Pmem_final}). The correction is up to 4.3\% percent at maximum. Galaxies with a photometric redshift accuracy $\sigma_0=0.03$ are considered.}
\label{fig:xi}
\end{figure}

{ 
The membership probabilities reported in Equations~\ref{eq:pmem5} and \ref{eq:pmem5tris} are { relative probabilities} in the sense that they refer to the chance a galaxy has of occupying an optimal (and relatively small) region  in the space of parameters that fiducial cluster members are associated with. In our case the adopted parameters are the redshifts, the cluster centric distance, and the {\sf H}-band magnitude of galaxies.
For example a galaxy in the cluster core with a photometric redshift close to that of the cluster will be associated with higher probability than a galaxy located at a different redshift and at the periphery of the cluster.

To illustrate better the concept we analytically compute the membership probability under the approximation that the PDFs $P_c(z)$ and $P_g(z)$ are both Gaussian functions, which is a few-percent precision approximation (see also Figure~\ref{fig:PDF_residuals}). It holds:
\small
\begin{align}
 \label{eq:Pmem_approx}
\mathcal{P}(g\in c|\Pi) = (1-\beta)\delta z (P_g\ast P_c)(z_c-z_g) = \nonumber\\
4.7\%\cdot(1-\beta) \Bigl(\frac{\delta z}{0.01}\Bigr)\Bigl(\frac{1+z_c}{2}\Bigr)^{-1}\Bigl(\frac{\sigma_0}{0.03}\Bigr)^{-1}e^{-\frac{(z_c-z_g)^2}{4(1+z_c)^2\sigma_0^2}}\,,
\end{align}
\normalsize
here $P_g\ast P_c$ denotes the convolution between the functions $P_g$ and $P_c$.

Our result implies that membership probabilities on the order of percent are expected for the cluster members.} This is not surprising: given the high photometric redshift uncertainties associated with galaxies ($\sim$~few 10~Mpc), if compared to the cluster physical size ($\sim1$~Mpc), the probability that both the cluster and a cluster member are located at the same distance from the observer along the line of sight is just on the order of $\sim1\%$ at maximum. { Furthermore the maximum probabilities are reached in the limiting case of negligible background or, equivalently, when the cluster is extremely rich (i.e., $\beta\ll1$). 

We note that, consistently with our definition, increasing the number of relevant parameters 
will increase the dimensionality of the parameter space thus reducing probabilities.

The membership probabilities are also not completely model independent quantities since they scale with the redshift bin $\delta z$. However, this dependence simply introduces the scaling $\mathcal{P}(g\in c|\Pi) \propto \delta z$ and can be neglected as far as  $\sigma_0(1+z_c) > \delta z \gg \delta z(g,c)$, at it is in our case. 

Our membership probabilities scale also linearly with the number count excess $(1-\beta)$, similarly to previous work by \citet{rozo2009}, and show a self-similar Gaussian decay in redshift, at least under the the pure Gaussian approximation.
Our estimate also shows a clear dependence on the cluster redshift as well as on the statistical photometric redshift accuracy of the considered catalog.
In fact, since photometric redshift uncertainties increase
with increasing redshift, the maximum achievable
probability decreases with increasing redshift, as expressed by the  
$\mathcal{P}(g\in c|\Pi)\propto\frac{1}{\sigma_0(1+z_c)}$ scaling in Equation~(\ref{eq:Pmem_approx}).

All these scaling relations inspired us to translate our relative probabilities into absolute membership probabilities $P_{mem}$ assuming self-similarity among clusters at different redshifts: we require that in the limit of negligible background or, equivalently, when the cluster is extremely rich (i.e., $\beta\ll1$) the  maximum membership probability achievable is exactly one, independently of the cluster redshift. 
In fact, in the limit where $\beta\ll1$ almost all galaxies within the cluster radius and around the cluster redshift are cluster members.

An effective way to exploit such a limit is to rescale the {relative probability} with respect to its maximum as follows:
\begin{equation}
 \label{eq:Pmem_final}
 P_{mem} = \frac{(1-\beta)\sum_iP_g(z_i)P_c(z_i)}{4.7\%\cdot\xi(z_c)\bigl(\frac{\delta z}{0.01}\bigr)\bigl(\frac{1+z_c}{2}\bigr)^{-1}\bigl(\frac{\sigma_0}{0.03}\bigr)^{-1}}\;.
\end{equation}
This is our final expression for the membership probability  we will use throughout this work.
The numerator of the Equation is equal to the relative probability $\mathcal{P}(g\in c|\Pi)$ and the denominator is its maximum value, which is reached at the limit of $\beta\ll1$, as in Equation~(\ref{eq:Pmem_approx}). 
The function $\xi(z_c)$ is the correction needed when the Gaussian approximation of $P_g(z)$ is relaxed. We estimate the function $\xi$ numerically. 

In Figure~\ref{fig:xi} we report $1-\xi$ as a function of redshift.
{Such a correction is limited to a few percent and increases with decreasing redshifts down to $z=0.09$ below which the support of the Gaussian PDF  ($\sigma_0=0.03$ is assumed) is significantly truncated at $z=0$. This truncation implies that $1-\xi$ decreases at lower redshifts ($z<0.09$) with decreasing redshift.} 
 }

\section{Results: testing the membership assignments}\label{sec:results}
In this Section we present our results and quantify their robustness in terms of completeness and purity of our membership assignments. 

For each cluster $N_{\rm estimated}$ is the number of galaxies that are considered cluster members according to the membership assignments, $N_{\rm true}$ is the number of {true} cluster members, $N_{\rm interlopers}$ is the number of sources that are erroneously considered cluster members according to the membership assignments, and $N_{\rm missed}$ is the number of {true} cluster members that are not selected. The four number counts are related as follows: $N_{\rm true} = N_{\rm estimated} + N_{\rm missed} - N_{\rm interlopers}$. 

{ Such numbers are always estimated within the $r_{200}$ radius, unless a radial interval is specified.}
{ Furthermore, when neither a radial nor a magnitude range is specified, the number $N_{\rm true}$ refers to the cluster richness or, equivalently, to the halo occupation number, which is the number of galaxies brighter than a given limit (${\sf H}_\ast+1.5$ in this work)  contained in a halo of a given mass \citep{Peacock_Smith2000}.}

Following the notation reported in \citet{george2011} we define purity \textsf{p} and completeness \textsf{c} of our assignments as:
\begin{equation}
 \mathsf{p}= 1-\frac{N_{\rm interlopers}}{N_{\rm estimated}}\,,
\end{equation}
\begin{equation}
 \mathsf{c}= 1-\frac{N_{\rm missed}}{N_{\rm true}}\,.
\end{equation}
Purity and completeness are related according to the following relation:
\begin{equation}
\label{eq:pc_Ntrue}
 N_{\rm true}=\frac{\mathsf{p}}{\mathsf{c}}N_{\rm estimated}\,.
\end{equation}
Such a relation is a powerful tool to estimate the true richness of the cluster ($N_{\rm true}$), once the number of selected cluster members ($N_{\rm estimated}$) and the ratio $\textsf{p}/\textsf{c}$ of our assignments are known.

\subsection{Completeness vs. Purity diagrams}\label{sec:cp_plot}

A practical way to select the fiducial cluster members on the basis of our membership assignments is to consider as cluster members all $N_{\rm estimated}$ galaxies that are associated with membership probabilities $P_{mem}$ higher than a given threshold $P_{thr}$ \citep[similarly to][]{george2011}, so that both purity and completeness can be parametrized by $P_{thr}$.
Such a prescription implies that the robustness of our assignments can be evaluated in terms of the Receiver Operating Characteristic \citep[ROC,][]{Metz1988}.

\begin{figure}[htbp]
\includegraphics[width=0.5\textwidth]{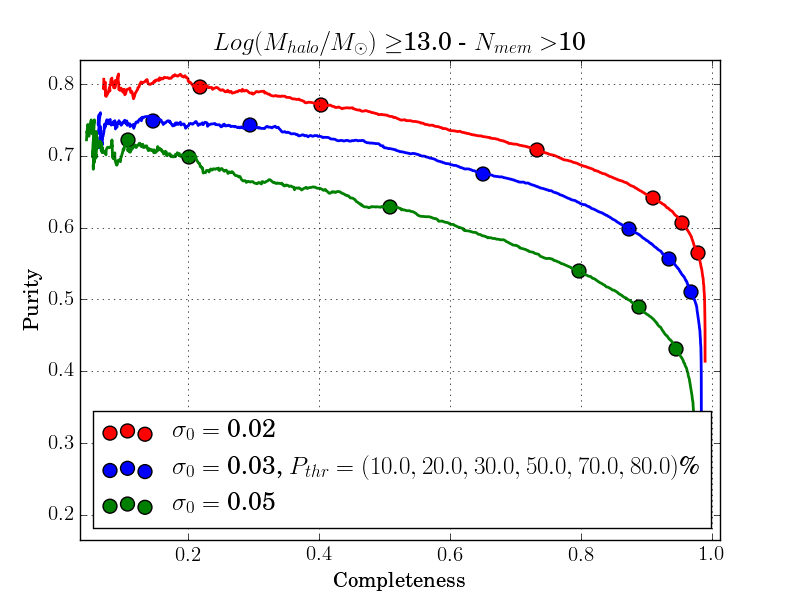}
\caption{\small Purity vs. completeness mean values for the membership assignments. {For each cluster galaxies brighter than ${\sf H}_\ast(z_{p})+1.5$ and 
with a projected distance from the cluster center not greater than $r_{200}$ are considered}. Different colors refer to different statistical redshift accuracy $\sigma(z)=\sigma_0(1+z)$. Dots show the mean values of both completeness and purity for galaxies with $P_{mem}>P_{thr}$, as indicated in the label.
$P_{thr}$ increases from the right to the left.
The errors in the mean values are within the dot size.}
\label{fig:pc_curve}
\end{figure}

In Figure~\ref{fig:pc_curve} we show the ROC curve associated with our assignments where all halos in our sample are considered. Mean values of both completeness and purity are plotted as a function of $P_{thr}$. The uncertainties in the mean values are at sub-percent level and therefore negligible. Concerning the photometric redshift catalog with $\sigma_0=0.03$ the mean values of purity and completeness are $\overline{\mathsf{p}}={0.51,  0.56,  0.60,  0.67,  0.74,  0.75}$
and  $\overline{\mathsf{c}}={0.97,  0.93,  0.87,  0.65,  0.29,  0.15}$, respectively. The different values refer to increasing $P_{thr}={10, 20, 30, 50, 70}$, and { 80}$\%$, respectively.
The corresponding median values are ${\sf p}_{median}={0.50^{+0.17}_{-0.14}}$,~${0.55^{+0.17}_{-0.15}}$,~${0.59^{+0.19}_{-0.17}}$, ${0.69^{+0.19}_{-0.21}}$, ${0.80^{+0.20}_{-0.30}}$, ${1.00^{+0.00}_{-0.67}}$ and ${\sf c}_{median}=
{1.00^{+0.00}_{-0.08}}$,~${0.95^{+0.05}_{-0.10}}$,~${0.90^{+0.10}_{-0.13}}$, ${0.68^{+0.14}_{-0.21}}$, ${0.28^{+0.22}_{-0.19}}$, ${0.10^{+0.17}_{-0.05}}$, where the uncertainties refer to the $68\%$ quartiles.

We then fix a threshold $P_{thr}={20\%}$, which allows us to have good mean values of both purity and completeness ($\overline{\mathsf{p}}={0.56}$ and $\overline{\mathsf{c}}={0.93}$), as well as a limited associated 68$\%$ dispersion  (${\sf p}_{median}={0.55^{+0.17}_{-0.15}}$ and ${\sf c}_{median}={0.95^{+0.05}_{-0.10}}$), which ultimately reflects into reliable richness estimates, see also Equation~(\ref{eq:pc_Ntrue}) and Section~\ref{sec:richness_estimate}.

Because of the non-negligible $68\%$ scatter, our membership assignments should be considered cautiously if used for single cluster studies. However given the good values of the mean for both purity and completeness as well as the negligible uncertainty in the mean we argue that the assignments are statistically robust, when large samples of clusters are considered.

\begin{figure*} \centering
\subfloat[]{\includegraphics[width=0.4\textwidth]{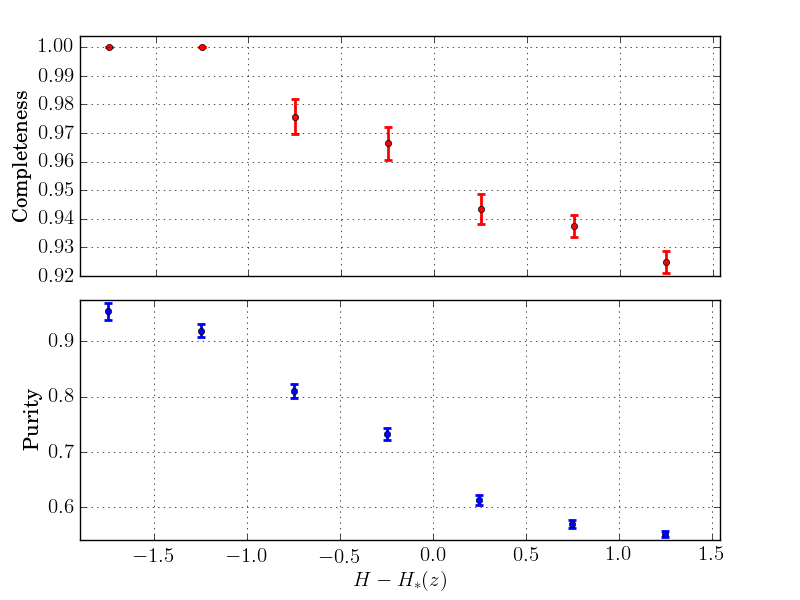}}\qquad
\subfloat[]{\includegraphics[width=0.4\textwidth]{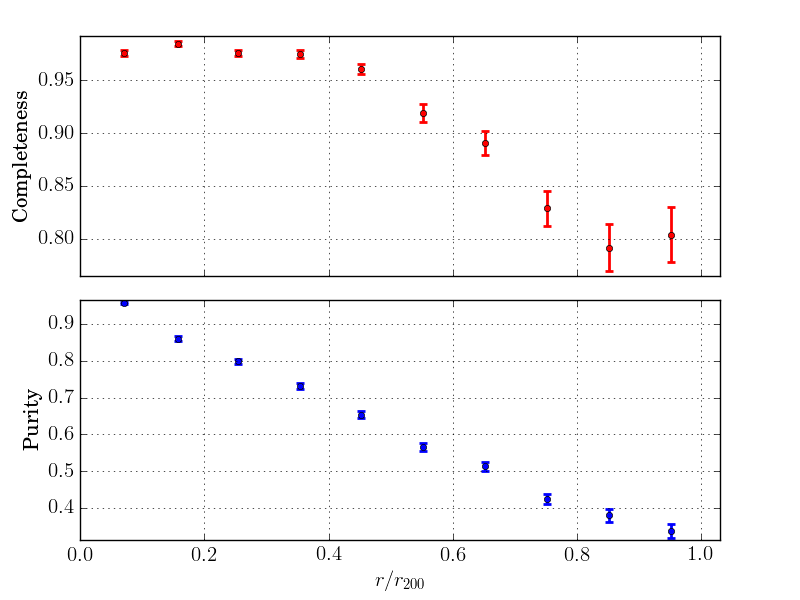}}\qquad
\subfloat[]{\includegraphics[width=0.4\textwidth]{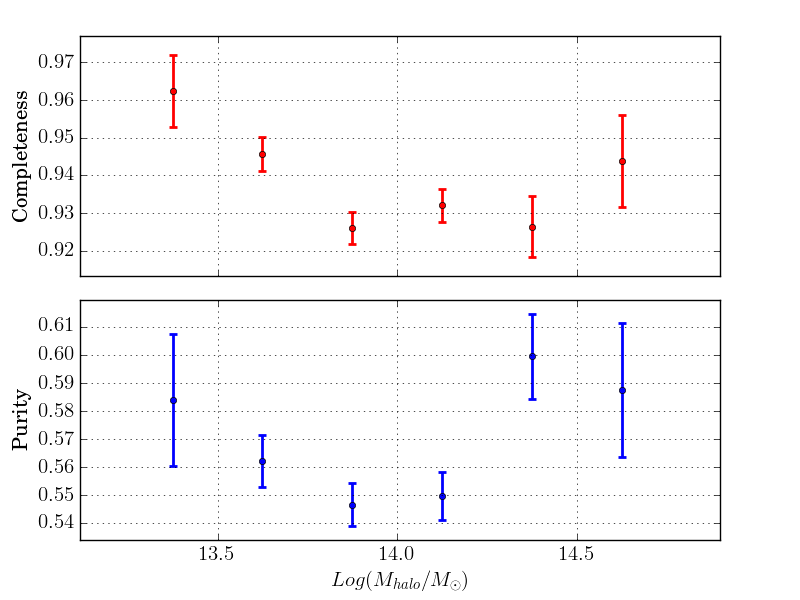}}\qquad
\subfloat[]{\includegraphics[width=0.4\textwidth]{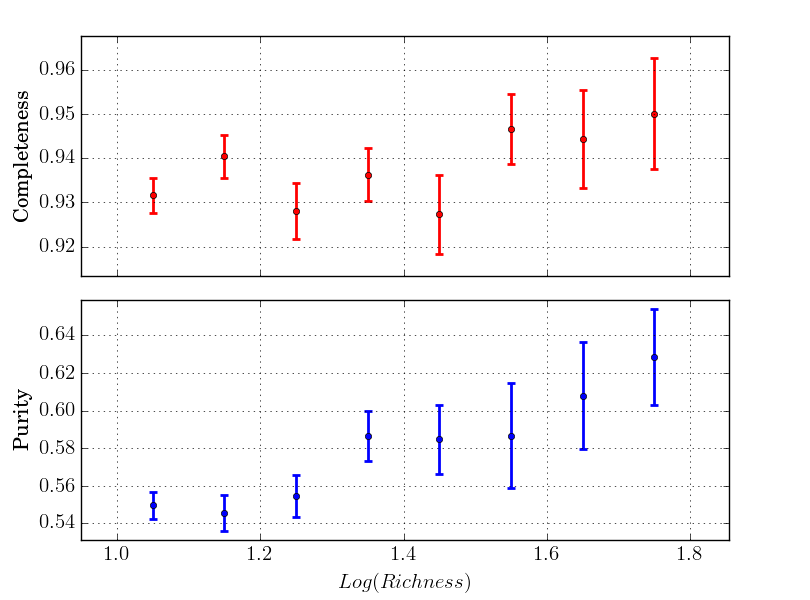}}\qquad
\subfloat[]{\includegraphics[width=0.4\textwidth]{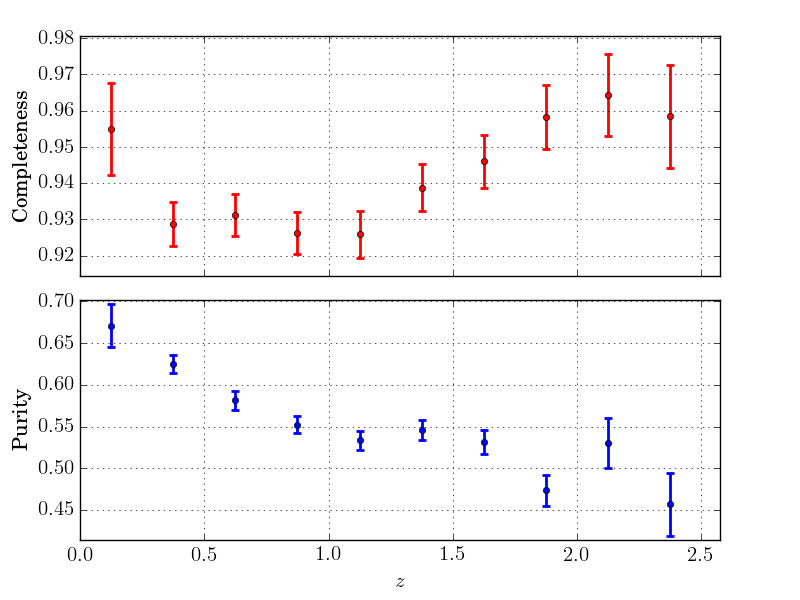}}\qquad
\subfloat[]{\includegraphics[width=0.4\textwidth]{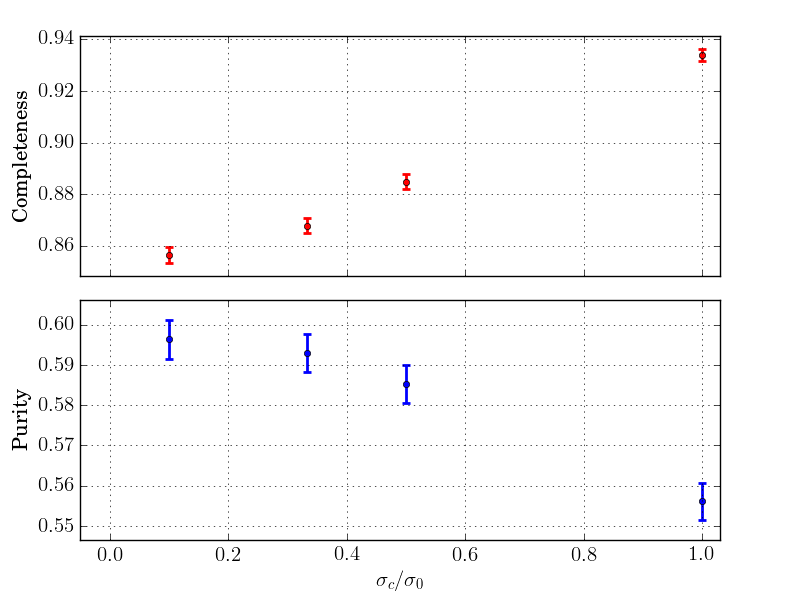}}\qquad
\caption{Mean completeness and purity along with associated errors as a function of galaxy magnitude (panel~a), projected separation, in $r_{200}$ units,  of the galaxy from the cluster center (panel~b), halo mass (panel~c) and richness (panel~d), cluster redshift (panel~e), and cluster redshift accuracy $\sigma_c/\sigma_0$ (panel~f). Richness refers to cluster galaxies brighter than ${\sf H}_\ast(z_{p})+1.5$ and within $r_{200}$ radius of each halo. All values refer to $\sigma_0=0.03$ and a probability threshold  $P_{thr}={20}\%$.}
\label{fig:cp_multi_panels}
\end{figure*}

In Figure~\ref{fig:cp_multi_panels} we show, for fixed $P_{thr}={20}\%$, the dependence of both completeness and purity on galaxy magnitude, separation of the galaxy from the cluster center, halo mass and richness, cluster redshift, and cluster redshift accuracy $\sigma_c$.

We observe remarkably stable mean values for both purity and completeness, with variations on the order of a few percent when the dependence on halo mass, richness, and cluster redshift accuracy are considered. A strong {decline}, in particular for purity is observed when faint  $\sim0.25\,L_\ast$ sources
($\Delta \mathsf{p}\simeq30\%$), and those at the outskirts of the halos ($\Delta \mathsf{p}\simeq50\%$, at cluster-centric projected distances $\sim r_{200}$) are considered, for which the contamination of field galaxies is significant.
Similarly a  {$\Delta \mathsf{p}\simeq20\%$} decline in purity is observed within the redshift range $z\sim0-2$.

\subsection{Membership probability distribution}\label{sec:Pmem_dist}
\begin{figure}[htbp]
\includegraphics[width=0.5\textwidth]{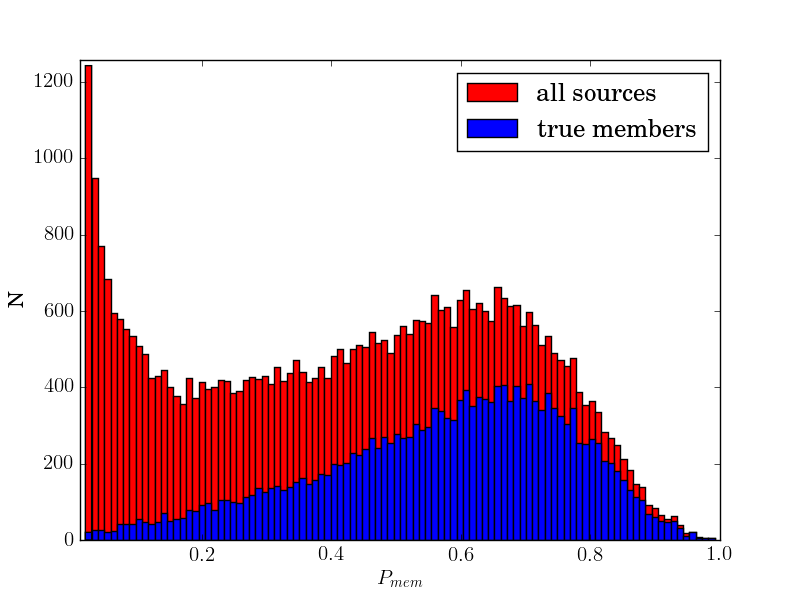}
\caption{\small Red: Distribution of membership probabilities for the galaxies with $P_{mem}>20\%$, brighter than ${\sf H}_\ast(z_{p})+1.5$, and {with a projected distance from the cluster center not greater than $r_{200}$.}
Galaxies with $P_{mem}>2\%$ are considered to avoid field galaxies associated with small membership probabilities.
Blue: same distribution but for the subsample of true cluster members.}
\label{fig:Pmem_histo}
\end{figure}

{ In Figure~\ref{fig:Pmem_histo} we show the distribution of the membership probabilities for all galaxies in the fields of the clusters in our sample, {i.e., with a cluster-centric projected distance not greater than $r_{200}$.} 
The median values of $P_{mem}$ for the galaxies with $P_{mem}>P_{thr} = 10, 20, 30, 50, 70$,  and 80$\%$ are $P_{mem,median}= 0.51\pm0.21, 0.55\pm0.18, 0.58\pm0.15, 0.66\pm0.10, 0.77\pm0.06$, and $0.84\pm0.04$, respectively. The reported uncertainty is the rms dispersion. 
These median values are consistent within a few percent with the corresponding mean values of purity $\overline{\mathsf{p}}={0.51,  0.56,  0.60,  0.67,  0.74,{\rm and\,}0.75}$ reported above, respectively, with the only exception represented by the last value associated with $P_{mem}>P_{thr} = 80\%$, for which marginal agreement is found.

The fact that the median values of $P_{mem}$ are both higher than the corresponding $P_{thr}$ and consistent with the associated $\overline{\mathsf{p}}$ ultimately explains the apparent discrepancy between 
$\overline{\mathsf{p}}$ and the corresponding $P_{thr}$ in the completeness vs. purity curve of Figure~\ref{fig:pc_curve}.

Furthermore, the distribution of Figure~\ref{fig:Pmem_histo} has a minimum around $P_{mem}\approx20\%$. The presence of such a minimum is mainly due to the contamination of field galaxies associated with relatively low membership probabilities.
The presence of the minimum is also due to the absence of faint galaxies with ${\sf H}>{\sf H}_\ast(z)+1.5$ and galaxies with a cluster-centric projected distance higher than the cluster radius, which are in fact not considered. They would nevertheless be associated with low values of $1-\beta$, thus populating the low-probability tail of the $P_{mem}$ distribution.

The location of the minimum at $P_{mem}\approx20\%$ also strengthens the choice of 
$P_{thr}=20\%$ as fiducial threshold (as for example in Figure~\ref{fig:cp_multi_panels}), limiting the contamination of field galaxies.
Furthermore, as mentioned above, galaxies with $P_{mem}>P_{thr} = 20 \%$ have $P_{mem,median}= 0.55\pm0.18$, which is consistent, within the dispersion, with analogous $P_{mem}>50\%$ cut adopted by previous work \citep{george2011}.}

\subsection{Fraction of true members}\label{sec:fraction_of_members}
{ 
In this Section we further test the robustness of our membership assignments. 
In Figure~\ref{fig:Fraction_of_members} we plot the fraction of true members ($f_{\rm true}$) for galaxies belonging to different bins of $P_{mem}$. We also consider four different redshift ranges.
An overall agreement between $P_{mem}$ and $f_{\rm true}$ is found, independently of the photometric redshift accuracy $\sigma_0$, within a few-percent precision, as reported in the Figure.

Figure~\ref{fig:Fraction_of_members} shows a flattening of $f_{\rm true}$ with increasing redshift for values $P_{mem}\gtrsim60\%$, where the membership probabilities are biased high, although at these $P_{mem}$ values $f_{true}$ is endowed with large uncertainty due to small number counts. 


We investigate statistically the goodness of our assignments evaluating the consistency of the $f_{\rm true}$ vs. $P_{mem}$ scatter plot with respect to the one-to-one line by means of $\chi^2$ statistics \citep[similarly to][see their Equations~36 and 37]{rozo2015}. 
Concerning the fraction $f_{\rm true}$, Poisson number count uncertainty is added in quadrature to a fiducial {(relatively small)} error $\delta f_{\rm true} = 3\%$. 

The $\delta f_{\rm true}$ correction is on the order of $f_{\rm true}$ vs. $P_{mem}$ offset and scatter reported in Figure~\ref{fig:Fraction_of_members} and outlined in the following. 
{The correction is effective in absorbing the $P_{mem}$ uncertainties and is ultimately needed to obtain reasonable reduced chi-square values $\chi^2_\nu\simeq1$
\citep[see e.g.,][for a similar approach in a different context]{Bourdin2015,Castignani_DeZotti2015}. 

In Table~\ref{tab:fraction_of_members} we summarize our results for different redshift bins. A few-percent $f_{\rm true}$ vs. $P_{mem}$ offset, i.e., $\langle f_{\rm true} - P_{mem}\rangle<0$, also shown in Figure~\ref{fig:Fraction_of_members}, is found. It is mainly due to the above mentioned flattening of $f_{\rm true}$ for high values of $P_{mem}$. In the following we further reconsider the scatter plot to understand the origin of the few-percent $f_{\rm true}$ vs. $P_{mem}$ discrepancy.



\begin{table}[htbc]
\centering
\begin{tabular}{ccccc}
\hline\hline
$\langle f_{\rm true} - P_{mem}\rangle$ &  $\chi^2_\nu$ &  d.o.f. &   $N_{halos}$ & $z$  \\
(1) & (2) & (3) & (4) & (5) \\
\hline
 $(+2.2\pm4.6)\%$ & $1.29$  & 10       &   179   & 0.0-0.5\\
 $(-2.1\pm5.2)\%$ & $0.97$  & 10       &   175   & 0.5-0.75\\
 $(-4.3\pm6.8)\%$ & $1.07$  & 10       &   245   & 0.75-1.0\\
 $(-8.6\pm10)\%$  & $4.68$  & 10       &   229   & 1.0-1.25\\
 $(-4.5\pm7)\%$   & $0.59$  & 10       &   159   & 1.25-1.5\\
 $(-9.4\pm14)\%$  & $1.97$  & 10       &   127   & 1.5-1.75\\
 $(-15\pm15)\%$   & $4.76$  & 10       &   60    & 1.75-2.0\\
 $(-5.7\pm6.1)\%$ &  $0.86$ & 9        &   33    & 2.0-2.5\\
 \hline
\end{tabular}
\caption{Membership results for rescaled probabilities $P_{mem}$. Column description. (1) mean value of the difference $f_{\rm true} - P_{mem}$ and rms uncertainty around the mean; (2) reduced $\chi^2$ estimated as $\chi^2_\nu=\chi^2/$d.o.f.; (3) degrees of freedom (number of $P_{mem}$ bins considered); (4) number of halos; (5) redshift bin.}\label{tab:fraction_of_members}
\end{table}

Faint and bright galaxies with magnitudes higher and lower than ${\sf H}_\ast(z)$ have  $\langle f_{\rm true} - P_{mem}\rangle = (-2.9\pm3.3)\%$ and $(-4.9\pm8.8)\%$, respectively. 
Similarly, rich and poor clusters with more and less than 25 members have $\langle f_{\rm true} - P_{mem}\rangle = (-0.8\pm3.3)\%$ and $(-5.7\pm7.9)\%$, respectively. 
The reported values agree with each other within the uncertainties at a few-percent level.

On the other hand a significant discrepancy of a few $10\%$ occurs when galaxies within and beyond $r_{200}/2$ are considered separately. Mean values  $\langle f_{\rm true} - P_{mem}\rangle = (+11\pm13)\%$ and $(-28\pm20)\%$ are in fact found, respectively.

In Figure~\ref{fig:Fraction_of_members_r_in_out} we show the scatter plot in the two cases and in the case where no radial cut is applied. In the last case an overall agreement with respect to the one-to-one line is found $\langle f_{\rm true} - P_{mem}\rangle = (-4.1\pm5.6)\%$, although a few-percent $f_{\rm true}$ vs. $P_{mem}$ discrepancy is still detected, on average. In the Figure, at variance with what has been done before, all galaxies and halos in the sample are altogether considered without dividing them in redshift bins.

We conclude that the discrepancy is originated neither by faint galaxies nor by poor clusters, but it is mainly due to the contamination of field galaxies at the outskirts of the clusters.
Such results are consistent with a significant few $10\%$ decrease $\Delta \mathsf{p}$ in purity with increasing cluster-centric projected distance reported in Figure~\ref{fig:cp_multi_panels}, also found in previous work \citep{george2011}.
Nevertheless, we stress that our sample is mainly constituted by small clusters and rich groups, for which 
an accurate consideration of the cluster profile is challenging. 
Such a difficulty could be overcome considering more massive clusters (work in preparation), where the contamination from field galaxies is less significant and/or including in our formalism the cluster radial profile (when known with sufficient accuracy) as additional prior information. 
In fact we checked that when we limit ourselves to the subsample of 27 clusters with a richness $N_{\rm true}\geq40$ our results significantly improve ($\chi^2/{\rm d.o.f.}=2.92/10$ and $\langle f_{\rm true} - P_{mem}\rangle = (1.7\pm2.0)\%$), which is consistent with the increase $\Delta \mathsf{p}\gtrsim10\%$ in purity and a few-percent increase in completeness with increasing richness, reported in Figure~\ref{fig:cp_multi_panels} within the richness range spanned by our sample.}

\begin{figure*}[htbp]
\subfloat{\includegraphics[width=0.4\textwidth]{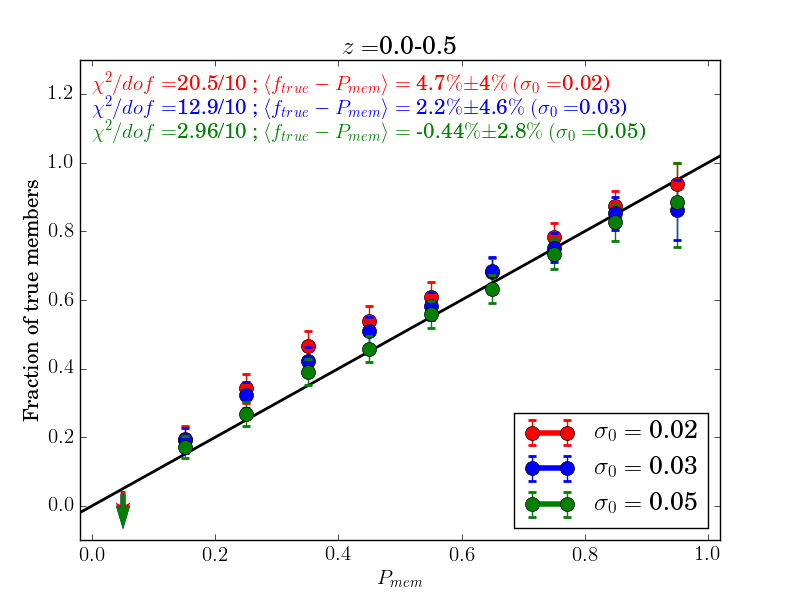}}
\subfloat{\includegraphics[width=0.4\textwidth]{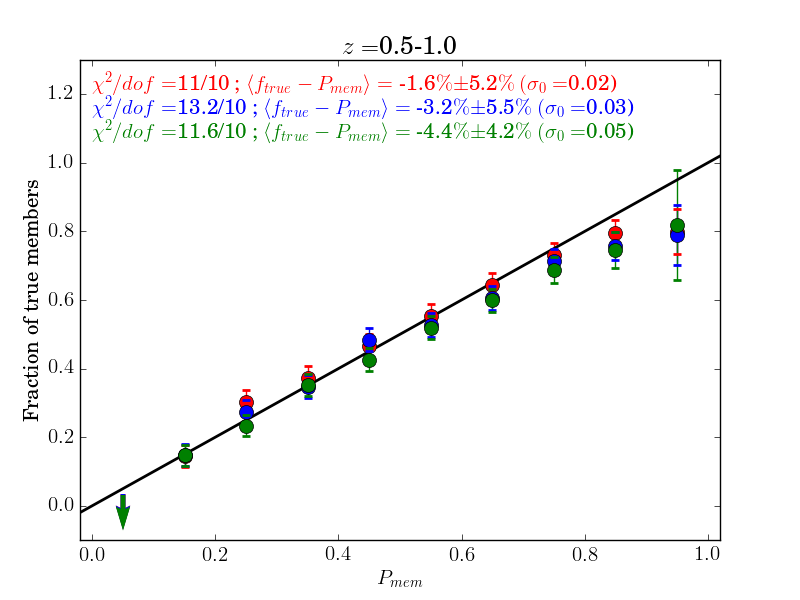}}\qquad
\subfloat{\includegraphics[width=0.4\textwidth]{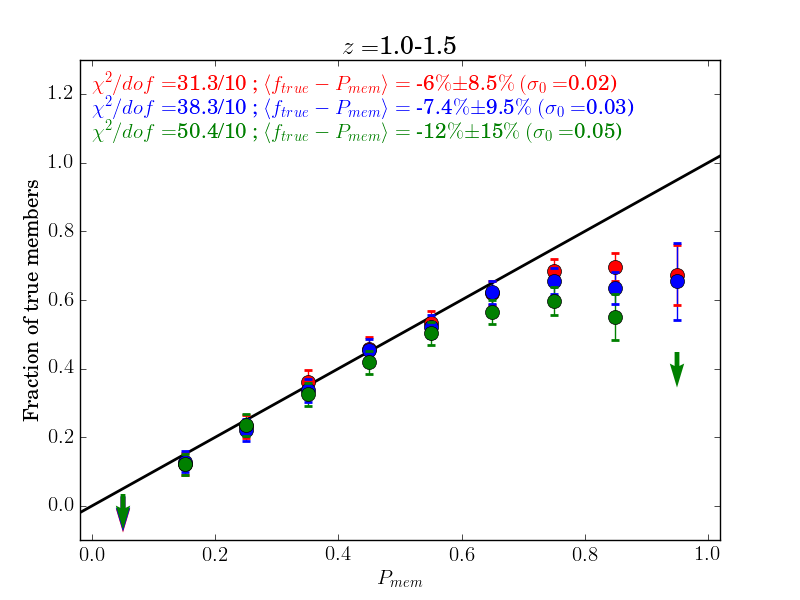}}
\subfloat{\includegraphics[width=0.4\textwidth]{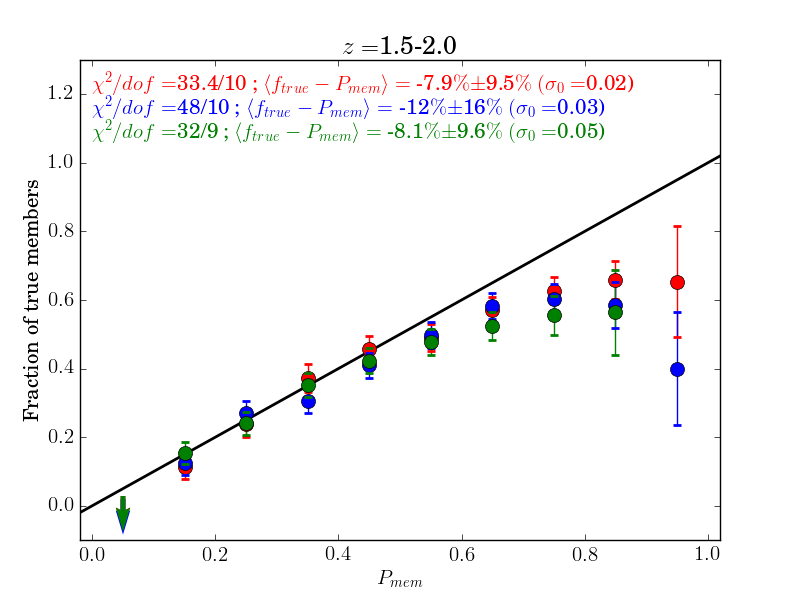}}\qquad
\caption{\small Fraction of true cluster members ($f_{true}$, y-axis) for galaxies brighter than ${\sf H}_\ast(z_{p})+1.5$ which membership probabilities ($P_{mem}$) reported in the x-axis are assigned to.
Points are associated with at least five sources per bins.
See Legend in the top left panel for the color code adopted. Different colors refer to different statistical redshift accuracy $\sigma(z)=\sigma_0(1+z)$. Mean values and Poisson uncertainties added in quadrature to $\delta f_{\rm true}=3\%$ are reported. Upper limits are at 2-$\sigma$ level.
Different panels refer to different redshift intervals. 
At the top of each panel the reduced $\chi^2$, the mean difference $\langle f_{\rm true} - P_{mem}\rangle$,  and the rms dispersion around the mean are reported. Upper limits are considered as true measurements when estimating $\langle f_{\rm true} - P_{mem}\rangle$.}
\label{fig:Fraction_of_members}
\end{figure*}

\begin{figure*}[htbp]
\subfloat[]{\includegraphics[width=0.4\textwidth]{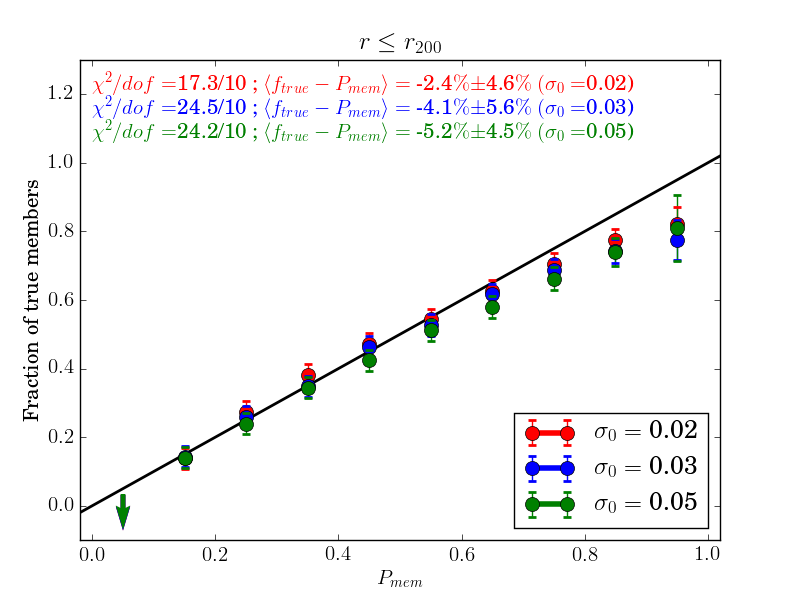}}
\subfloat[]{\includegraphics[width=0.4\textwidth]{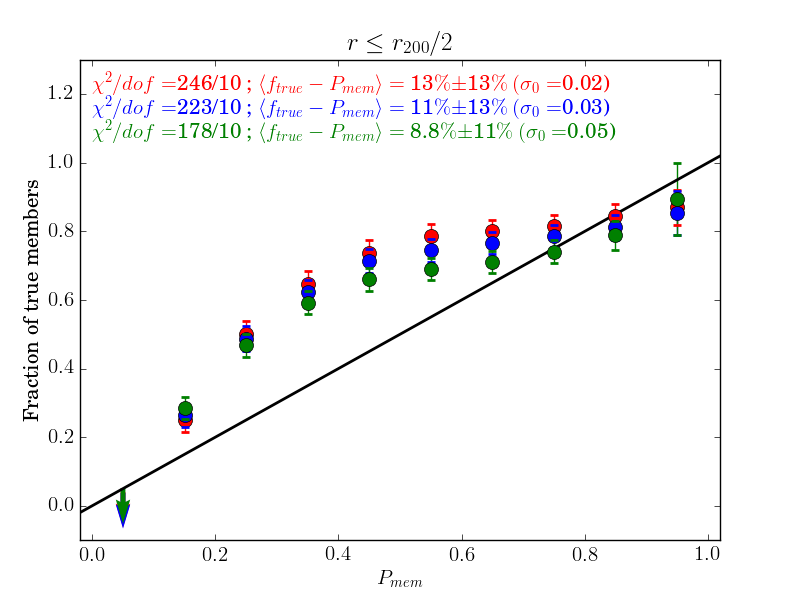}}\qquad
\subfloat[]{\includegraphics[width=0.4\textwidth]{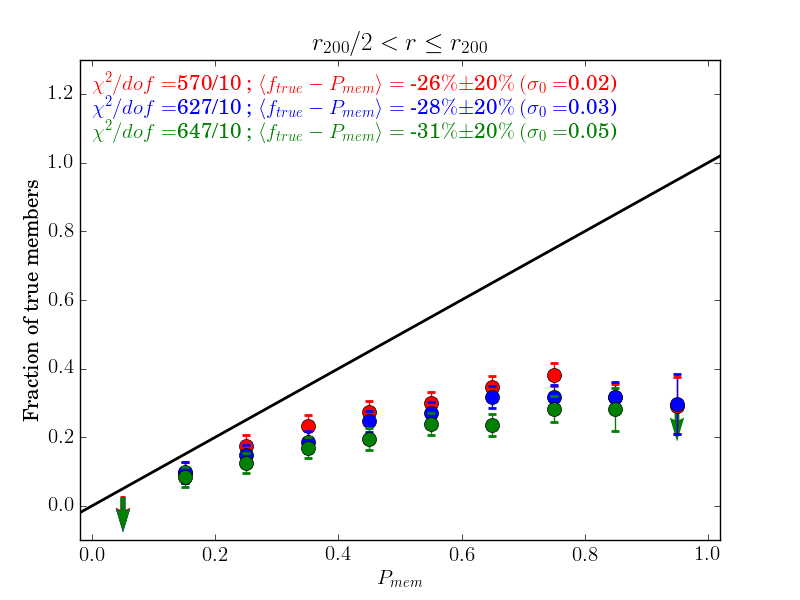}}\qquad
\caption{\small Fraction of true cluster members vs. $P_{mem}$, as in Figure~\ref{fig:Fraction_of_members}, for all halos in the sample.
Panel (a) refers to galaxies within $r_{200}$.
Panels (b) and (c) refer to galaxies within and outside $r_{200}/2$, respectively. Color code is the same for all panels.}
\label{fig:Fraction_of_members_r_in_out}
\end{figure*}

\subsection{Correlated structures}

{ In this Section we examine how the presence of correlated structures affects our results.
First we exploit our simulations rejecting from both local and global background areas, when estimating the background, all sources belonging to halos with masses $\geq10^{13}~M_\odot$. This has the net effect of increasing $P_{mem}$, thus increasing also both the average bias $\langle f_{\rm true} - P_{mem}\rangle=(-6.6\pm4.6)\%$ and $\chi^2/{\rm d.o.f.}=47.4/10$ with respect to the case where no correlated structure is removed and all halos in the sample are considered.

When galaxies belonging to halos more massive than $10^{13}~M_\odot$ are rejected also from the cluster field, {i.e., a projected distance from the cluster center not greater than $r_{200}$ is considered}, the average bias disappears ($\langle f_{\rm true} - P_{mem}\rangle=(-0.87\pm9.1)\%$).
{However both the scatter ($9.1\%$) and associated $\chi^2/{\rm d.o.f.}=47.8/10$ are higher than the values $5.6\%$ and  
$\chi^2/{\rm d.o.f.}=24.5/10$, respectively, obtained in the case where no correlated structure is removed (see Section~\ref{sec:fraction_of_members}).} 
This is ultimately due to the fact that when correlated structures are removed from the cluster field number counts are reduced and higher shot-noise fluctuations affect the results implying the observed higher scatter and higher reduced chi-square values.

{Therefore, we find that correlated structures affect the observed $P_{mem}$ vs. $f_{true}$ offset similarly (but less dramatically) to field galaxies at the outskirts of the cluster (see Section~\ref{sec:fraction_of_members}). Our results also suggest that correlated structures have to be removed both in the cluster field and in the background areas to reduce the $P_{mem}$ bias. Nevertheless this approach is critical and does not ultimately improve the results in terms of $P_{mem}$ vs. $f_{true}$ scatter, on average, in our case of low-number statistics and relatively poor clusters.}


Interestingly, similarly to our results, \citet{rozo2015} found that their membership probabilities are biased high {when correlated structures are not removed \citep[see also][]{rykoff2012,rykoff2014}}. They corrected the probabilities for the presence of correlated structures, assuming that the number of galaxies belonging to them is a constant fraction of the cluster richness.
We stress nevertheless that their sample is mainly constituted by rich clusters, at variance with our sample, for which low-number statistics is affecting more the number counts.
Therefore, we prefer not to apply any correction, which implies that we avoid specific assumptions on the properties of correlated structures, in agreement with our non-parametric formalism.


}

\subsection{Comparison with previous work}
{ In this Section we mainly focus on the comparison of our results with those of \citet{george2011}, who used photometric redshift information to address membership for clusters up to $z=1$ and halo masses $M_{halo}\lesssim10^{14}~M_\odot$. {They use both real photometric redshifts and simulated galaxy mock catalogs with Gaussian photometric redshifts. The latter case is analogous to ours.}}

As further outlined in the following our results are globally consistent with those of \citet[][]{george2011} even if 
{ better completeness and slightly better purity seem to be achieved in our work.} 
We stress that a strict comparison with previous work is nevertheless impossible mainly because of the different dataset used (e.g., different  photometric redshift uncertainties are considered) and the different range of halo masses and redshifts considered.

{ {Limiting ourselves to the 2~square degree COSMOS survey \citep{scoville2008} and real photometric redshifts} the purity reported in \citet{george2011} tends to saturate at high membership probabilities to values $\sim80\%$ that are similar to ours for $\sigma_0=0.02$ (see their Figure~4). This occurs despite the small photometric redshift accuracy $\sigma(z)=(0.01-0.02)(1+z)$ of the COSMOS survey \citep{ilbert2009}.}

{ {Considering simulated galaxy mock catalogs and in particular that with associated simulated Gaussian photometric redshifts}} with an accuracy $\sigma_0=0.05$, they report values $\overline{\mathsf{p}}\simeq(50-55)\%$ and {$\overline{\mathsf {c}}\lesssim70\%$} (their Figure~7).
Conversely, as shown in Figure~\ref{fig:pc_curve} for $\sigma_0=0.05$, we have $\overline{\mathsf{p}}={49}\%$ (${\mathsf{p}}_{median}={0.47^{+0.17}_{-0.15}}$) and $\overline{\mathsf{c}}={89}\%$ (${\mathsf{ c}}_{median}={0.91^{+0.09}_{-0.11}}$) for $P_{thr}={20}\%$.\footnote{ The $P_{thr}={20}\%$ threshold  corresponds here (for $\sigma_0=0.05$) to galaxies with a median $P_{mem}=(48\pm17)\%$.} { Therefore similar purity values are found and significantly higher 
completeness ($\Delta \mathsf{c}\gtrsim20\%$) is remarkably found in our case.}

{ Moreover we checked that when we limit ourselves to the redshift range $z=0-1$, as in \citet{george2011}, the purity improves, $\Delta \mathsf{p}\simeq5\%$, strengthening our results.}

{ As it will be clarified in the following Section, we suggest that the differences with respect to previous studies {might be due} to a different consideration of the photometric redshift information when estimating the membership probabilities.
}

\subsection{Reconsidering the $P_{mem}$ rescaling}
{ As described in Section~\ref{sec:Pmem_rescaling} the {relative} membership probabilities are rescaled assuming self-similarity among clusters at different redshifts to obtain {absolute probabilities}, which reflect the actual values of purity reported in Sections~\ref{sec:cp_plot} and \ref{sec:fraction_of_members}, see also Figure~\ref{fig:Fraction_of_members}.

An alternative approach would be to define directly the membership probability in such a way that the associated values are close to one for fiducial cluster members, without the need of rescaling. This is a strategy which is commonly adopted in previous work \citep{brunner2000,george2011,rozo2015}.
Two possible approaches are exploited in the following.}

\subsubsection{Kernel $\phi$}
{We test, within our formalism, the impact of enlarging the support of the function $\phi(z_c',z_{s,g}')$, which is assumed to be a Kronecker delta, see Equations~(\ref{eq:pmem5},\ref{eq:pmem5tris}).  A support with a width  $|z_c'-z_{s,g}'|<2\sigma_0(1+z_c)$ is here adopted for $\phi$.
Enlarging the support of $\phi$ has the net effect of boosting the membership probabilities towards higher values for galaxies at redshifts close to that of the cluster. This approach is similar to that adopted in previous analyses \citep{brunner2000,rozo2015}.

Two different $\phi$ functions are tested here: a top-hat kernel $\phi_{top-hat}(z_c',z_{s,g}')\equiv1$ 
and a Gaussian kernel $\phi_{Gauss}(z_c',z_{s,g}')=e^{-\frac{(z_c'-z_{s,g}')^2}{2[\sigma_0(1+z_c)]^2}}$.

When all clusters in the sample are considered, if $\phi_{top-hat}$ and $\phi_{Gauss}$ are separately adopted we obtain, on average, $\langle f_{\rm true} - \mathcal{P}(g\in c|\Pi)\rangle = (-0.88\pm3.9)\%$ ($\chi^2/{\rm d.o.f.}=15.1/9=1.7$) and $\langle f_{\rm true} - \mathcal{P}(g\in c|\Pi)\rangle = (+15\pm9.9)\%$ ($\chi^2/{\rm d.o.f.}=166/6=27.7$), respectively, as opposed to $\langle f_{\rm true} - P_{mem}\rangle = (-4.1\pm5.6)\%$ ($\chi^2/{\rm d.o.f.}=24.5/10=2.5$), which is found in the case of rescaled membership probabilities (see Figure~\ref{fig:Fraction_of_members_r_in_out}a).

Biased low membership probabilities are found when the Gaussian kernel $\phi_{Gauss}$ is used, which in fact gives less weight to the wings than $\phi_{top-hat}$, within the 2-$\sigma(z)$ support. 

Results comparable to those obtained using rescaled probabilities are obtained when the top-hat filter $\phi_{top-hat}$ is adopted. A careful comparison of the results associated with the two cases when redshift bins are considered is reported in Table~\ref{tab:fraction_of_members_top_hat} and shows that 
rescaled probabilities $P_{mem}$ implies globally smaller $\chi^2_\nu$ values and are therefore preferred.


We also mention that in the case of rescaled probabilities we reach $P_{mem}\gtrsim90\%$ and similar $f_{\rm true}$ values, although endowed with uncertainties. On the contrary, when $\phi_{top-hat}$ and $\phi_{Gauss}$ are assumed, probabilities are  $55.7\%$ and $83.4\%$ at maximum, respectively. 

Our analysis also shows that, when no rescaling is applied, the results are strongly sensitive to the specific functional form of the kernel $\phi$ (here $\phi_{top-hat}$ and $\phi_{Gauss}$ are considered). Other, rather subjective, choices of the kernel would lead to different results. This last aspect strengthens our choice of $\phi$, which is  Kronecker delta, and the {\it a posteriori} rescaling of the probabilities which exploits self-similarity.

\begin{table*}[htbc]
{
\centering
\begin{tabular}{cccccc}
\hline\hline
$\langle f_{\rm true} - \mathcal{P}(g\in c|\Pi)\rangle$ &  $\chi^2_{\nu,top-hat}$ &  d.o.f.$_{top-hat}$ & $\chi^2_\nu$ &  d.o.f. &  $z$  \\
(1) & (2) & (3) & (4) & (5) & (6) \\
\hline
 $(+5.8\pm6.7)\%$ & $3.26$  & 9   & 1.29  &  10    & 0.0-0.5\\
 $(-0.1\pm4.1)\%$ & $1.00$  & 9   & 0.97  &  10     & 0.5-0.75\\
 $(-0.7\pm4.5)\%$ & $1.78$  & 8   & 1.07  &  10    & 0.75-1.0\\
 $(-6.5\pm7.2)\%$  & $3.43$  & 9  & 4.68  &  10     & 1.0-1.25\\
 $(-1.5\pm6.0)\%$   & $2.49$  & 8 & 0.59  &  10     & 1.25-1.5\\
 $(-4.4\pm5.7)\%$  & $2.08$  & 8  & 1.97  &  10    & 1.5-1.75\\
 $(-8.8\pm6.0)\%$   & $3.41$  & 8  & 4.76 &  10     & 1.75-2.0\\
 $(-4.1\pm4.6)\%$ &  $1.85$ & 8    & 0.86 &  9    & 2.0-2.5\\
 \hline
\end{tabular}
\caption{Membership results (columns 1-3) when the top-hat kernel $\phi_{top-hat}$ is adopted: (1) mean value of the difference $f_{\rm true} - \mathcal{P}(g\in c|\Pi)$ and rms uncertainty around the mean, upper limits of $f_{\rm true}$ at 2$\sigma$ level are considered as true measurements; (2) values of the reduced $\chi^2$ estimated as $\chi^2_\nu=\chi^2/$d.o.f.; (3) degrees of freedom (number of $\mathcal{P}(g\in c|\Pi)$ bins considered); (4) and (5) are analogous to columns 2 an 3 (see also Table~\ref{tab:fraction_of_members}), respectively, when rescaled probabilities are adopted;
(6) redshift bin.}\label{tab:fraction_of_members_top_hat}}
\end{table*}

In the next Section we exploit a second independent approach to obtain absolute probabilities without any rescaling.}

\subsubsection{PDF$(z)$ of galaxies in the cluster field}
{ Our formalism makes use of the full PDF$(z)$ of each galaxy, as reported in Equation~(\ref{eq:prob2}). 
We revisit the expression for the PDF evaluating $P_g(z_{s,g}|z_{c})$, which is needed in Equation~(\ref{eq:pmem1}). We write: 


\begin{equation}
\label{eq:PDFmod}
 P_g(z_{s,g}|z_{c}) = (1-\beta)\delta(z_{s,g}-z_c) +\beta P_g(z_{s,g})\,,
\end{equation}
which is a convex linear combination of a Dirac delta centered at the cluster redshift and the  PDF $P_g(z_{s,g})$ of Equation~(\ref{eq:prob2}). 
This approach is similar to that adopted in previous studies \citep[][]{george2011,rozo2015}.

The convex linear combination is such that, in the presence of a poor cluster, i.e., $\beta\lesssim1$, the PDF of the  galaxy is close to that of Equation~(\ref{eq:prob2}).
On the contrary, in the presence of rich clusters, i.e., $\beta\ll1$, the PDF of the galaxy is centered at the cluster redshift, independently of its photometric redshift.

Therefore the PDF of Equation~(\ref{eq:PDFmod}) operatively takes the clustering of photometric redshifts around the cluster redshift into account.
To reject conservatively field galaxies the PDF of Equation~(\ref{eq:PDFmod}) is assigned
only to galaxies within a $\pm2\sigma_0(1+z_c)$ interval centered at the cluster redshift $z_c$.
Then, we introduce such a PDF in Equation~(\ref{eq:pmem5tris}) to obtain a new estimate for the membership probability. Here we adopt the Kronecker delta as $\phi$ function. 
By definition, we note that, independently of the functional form of $\phi$, the membership probability now tends to unity in the limit of rich cluster ($\beta\ll1$), without any rescaling/modification of the membership probability. We checked that the results (outlined below) do not improve if we consider instead $\phi_{Gauss}$ or $\phi_{top-hat}$. \\

We observe a significant $\Delta \mathsf{p}\,,\Delta \mathsf{c}\sim5\%$ decrease of the maximum values achievable for both purity and completeness,  with respect to the case where rescaled probabilities are adopted, when averaging among all clusters in our sample. Similarly, we find  $\langle f_{\rm true} - \mathcal{P}(g \in c)\rangle = (2.8\pm10)\%$ and $\chi^2/{\rm d.o.f.} = 81/10$, when considering all galaxies in the sample.
The reduced $\chi^2$ is significantly higher than that obtained when rescaled probabilities are adopted ($\chi^2/{\rm d.o.f.} = 24.5/10$).\\

We conclude that our rescaled probabilities imply better results concerning both purity and completeness of our assignments with respect to the different tested cases, where i) the PDF$(z)$ of galaxies in the
field of clusters {(i.e., galaxies with a projected distance from the cluster center not greater than $r_{200}$}) is modified to take operatively the presence of the overdensity into account  and/or ii) the support of the function $\phi$ is enlarged. In both cases the maximum probabilities achievable have values close to unity, without the need of rescaling. 

However for both tested cases the broader range of redshift allowed for the selected cluster members implies higher contamination from non-cluster members (observed in terms of worse purity and higher $\chi^2_\nu$ values) than in the case of our rescaled probabilities, where fiducial cluster members are confined to an optimal region of the space of parameters.

}

\subsection{Local vs. global background}
We compare our results obtained with a local estimate of the background with those obtained using the global background.
Our results do not change, statistically, if the global background estimate is used instead of the local. This is ultimately due to the relatively small dispersion of the $f$-factor distribution reported in Figure~\ref{fig:f_factorhisto}. 
The results are unchanged even when the tails of the $f$-factor distribution are considered, as will be clarified in the following.

We consider two subsamples of 30 halos (each of the two corresponds to $2.5\%$ of the entire halo sample) associated with ${f<0.7522}$ and ${f>1.537}$, respectively. 
We remind that low (high) values of $f$ correspond to low (high) local-to-global background ratios.
In the following we show the results for fixed threshold $P_{thr}={20}\%$.

When $f<0.7522$ and the local background are considered the median (mean) completeness and purity are ${\mathsf{c}}_{median}={0.95^{+0.05}_{-0.04}}$ ($\overline{\mathsf{c}}={95}\%$) and ${\mathsf{p}}_{median}={0.59^{+0.18}_{-0.09}}$ ($\overline{\mathsf{p}}={61}\%$), respectively.
When the same threshold $f<0.7522$ and the global background are considered the completeness decreases and the purity increases, statistically. The median (mean) values are  ${\mathsf{c}}_{median}={0.92^{+0.07}_{-0.06}}$ ($\overline{\mathsf{c}}={91}\%$) and ${\mathsf{p}}_{median}={0.63^{+0.22}_{-0.12}}$ ($\overline{\mathsf{p}}={65}\%$), respectively.

When $f>1.537$ and the local background are considered the median (mean) completeness and purity are ${\mathsf{c}}_{median}={0.92^{+0.08}_{-0.08}}$ ($\overline{\mathsf{c}}={91}\%$) and ${\mathsf{p}}_{median}={0.53^{+0.21}_{-0.11}}$ ($\overline{\mathsf{p}}={49}\%$), respectively.
When the same threshold $f>1.537$ and the global background are considered the completeness increases and the purity decreases, with respect to the case of local background. { This corresponds to an opposite behavior with respect to the case when low values of $f$ are considered and it is a consequence of the fact that for low (high) values of $f$, going from local to global background implies an increase (decrease) of $N_{bkg}$ and a consequent decrease (increase) of the membership probability.}
The median (mean) values when the global background is considered and $f>1.537$ are in fact  ${\mathsf{c}}_{median}={0.98^{+0.02}_{-0.07}}$ ($\overline{\mathsf{c}}={95}\%$) and ${\mathsf{p}}_{median}={0.42^{+0.14}_{-0.11}}$ ($\overline{\mathsf{p}}={43}\%$), respectively.

Therefore, overall {$\Delta\mathsf{p}\,,\Delta\mathsf{c}\lesssim5\%$} variations in both purity and completeness are observed when considering one background estimate or the other.
A significant {decrease} of the purity $\Delta\mathsf{p}\sim10-20\%$ (nevertheless statistically consistent within the reported 68\% uncertainty)  is observed when going from low to high values of $f$. This is expected since the last case corresponds to halos where the local background is intrinsically high.\\

{ Interestingly, since $f$ is an observable quantity it could be used in the case of single cluster studies to adjust optimally the threshold $P_{thr}$ in order to achieve better purity, in particular for clusters with high local background.}

\section{Results: richness estimates}\label{sec:richness_estimate}
A remarkable byproduct of our assignments is that we can use them to perform richness estimates ($\lambda$). Two richness definitions are considered: i) $\lambda_1=N_{\rm estimated}$ \citep[][]{george2011}, which is the number of sources with $P_{mem}>P_{thr}$, where the probability threshold $P_{thr}={20}\%$ is applied, as described above and ii) $\lambda_2=\sum P_{mem}$ \citep{rozo2009}, where no probability threshold is applied and galaxies are weighted by their membership probability. 

\begin{figure*}[tbhc] \centering
\includegraphics[width=0.7\textwidth]{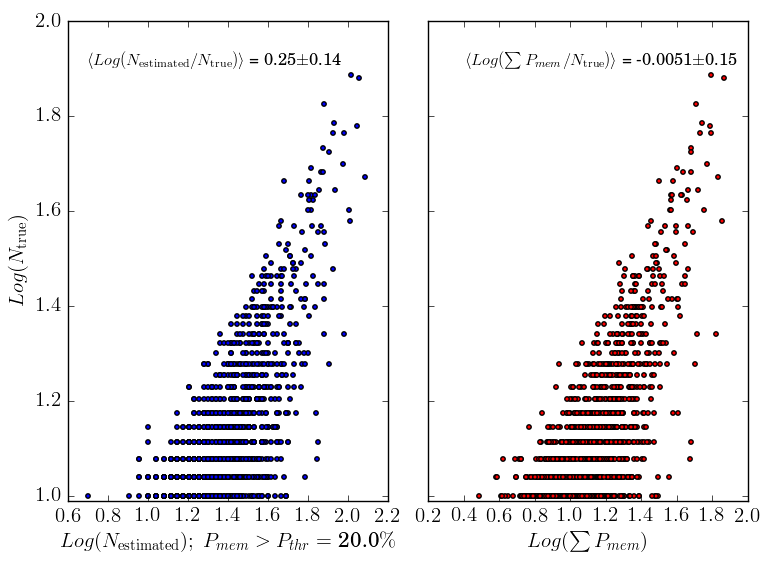}
\caption{True vs. estimated richness for the 1,208 halos in our sample. Richness refers to galaxies brighter than ${\sf H}_\ast(z_{p})+1.5$ and within $r_{200}$ radius of each halo. 
Galaxies with a statistical photometric redshift accuracy $\sigma_0=0.03$ are considered.
Left: richness is estimated as $N_{\rm estimated}$, where $P_{thr}={20\%}$ is considered. Right: richness is estimated as the sum of the membership probabilities. The mean logarithmic difference along with the rms dispersion around the mean are reported in the panels.}
\label{fig:richness}
\end{figure*}


\subsection{Comparing the two richness estimates}
When all halos in the sample are considered (Figure~\ref{fig:richness}) the mean logarithmic difference between the estimated ($\lambda$) and true richness  is
$\langle Log(N_{\rm estimated}/N_{\rm true}) \rangle={0.25\pm0.14}$ and $\langle Log(\sum P_{mem}/ N_{\rm true}) \rangle={-0.0051\pm0.15}$, where the reported uncertainty is the rms dispersion around the mean.\footnote{{ {We verified that the results only slightly change if the threshold 
$P_{thr}={20\%}$ is applied also when estimating the richness $\lambda_2$. In fact in this case we find $\langle Log[(\sum_{P_{mem}>P_{thr}} P_{mem})/ N_{\rm true}] \rangle=-0.029\pm0.16$.} }}
We apply the Spearman test. Evidence of clear correlation between $Log(N_{\rm true})$ and $Log(\lambda)$ is found (p-value $<$1e-150), independently of the richness estimate considered. The values reported in the present and following Sections refer to galaxies with a statistical photometric redshift accuracy $\sigma_0=0.03$. However we verified that the richness estimates are remarkably stable when a different photometric redshift accuracy is considered: $\langle Log(\sum P_{mem}/ N_{\rm true}) \rangle=-0.0064\pm0.14$ and $-0.0011\pm0.17$, for $\sigma_0=0.02$ and $0.05$, respectively.

Our results show that both adopted richness estimates $\lambda$ lead to similar results, in terms of rms dispersion. When the mean purity and completeness of the assignments are considered, i.e., $Log(\overline{\mathsf{p}}/\overline{\mathsf{c}})=-0.22$ (see Equation~\ref{eq:pc_Ntrue}), the mean offset between $\lambda_1=N_{\rm estimated}$ and $N_{\rm true}$ disappears, within the rms dispersion. 
Richness estimates derived as $\lambda_2=\sum P_{mem}$ are instead remarkably unbiased with respect to the true richness $N_{\rm true}$, on average.

In Tables~\ref{tab:richness_results} and \ref{tab:richness_results2} we report our results for different redshift and mass intervals, respectively.
{ Our results 
show that richness estimates are provided with a $\sim0.10-0.17$~dex accuracy, almost independently of the redshift bin considered, within a few 0.01~dex. 
The lowest dispersion, $\sim$0.07-0.08~dex, is reached for high  halo masses, $M_{halo}\geq 10^{14.5}~M_\odot$.}
Therefore, as clarified in the following, our results are fairly independent of the redshift and halo mass intervals considered.

\subsection{Richness as mass proxy}

{ We exploit our richness estimates as halo mass proxy. Estimating observable quantities which are tightly correlated with the cluster mass is in fact essential for cosmological studies. First we consider all halos in our sample. The rms dispersion around the mean of the logarithmic difference between the halo mass and the halo occupation number is $\sigma(Log(M_{\rm halo}/N_{\rm true}))=0.14$. Such a scatter is consistent with previous studies \citep{Zheng2005} and is related to the purely cosmological halo occupation distribution, i.e., the richness distribution for halos of fixed mass. Similarly, when considering our richness estimates we obtain $\sigma(Log(M_{\rm halo}/N_{\rm estimated}))=0.20$ and $\sigma(Log(M_{\rm halo}/\sum P_{mem}))=0.20$. Our results imply that, to recover the observed halo mass vs. observed richness scatter, an uncertainty of $\sim0.14$~dex, on average, has to be summed in quadrature to the intrinsic $\sigma(Log(M_{\rm halo}/N_{\rm true}))\sim0.14$ scatter, consistently with the relation $\sigma(Log(M_{\rm halo}/\sum P_{mem}))\simeq\sqrt{2}\sigma(Log(M_{\rm halo}/N_{\rm true}))$.}


{  
Then we consider different redshift and halo mass intervals, separately (see Tables~\ref{tab:richness_results} and \ref{tab:richness_results2}).
The richness vs. halo mass scatter varies within $\sim$0.10-0.21~dex, where the lowest values, 0.10 and 0.14~dex, are reached for halo masses $M_{halo}\geq 10^{14.5}~M_\odot$ and redshifts $z\leq0.75$, respectively.  These values are remarkably similar to those $\sim0.1-0.25$ found independently using scaling relations based on X-ray and  weak lensing studies \citep{Sereno_Ettori2015}, by studies of Sunyaev-Zel'dovich detected clusters \citep{Sereno2015}, as well as by other methods based on galaxy number counts \citep{andreon2015}. We stress that all these analyses are limited to massive, intermediate/low redshift, and typically relaxed clusters, while our sample is mainly constituted by rich $N_{\rm true}\geq10$ groups spanning a wide range in redshift. 
 }


\begin{table*}[htbc]
\centering
{\small
\begin{tabular}{cccccccc}
\hline\hline
$\langle Log(\frac{N_{\rm estimated}}{N_{\rm true}})\rangle$  & $\langle Log(\frac{\sum P_{mem}}{N_{\rm true}}) \rangle$  & $\sigma(Log(\frac{M_{\rm halo}}{N_{\rm true}}))$ & $\sigma(Log(\frac{M_{\rm halo}}{N_{\rm estimated}}))$ & $\sigma(Log(\frac{M_{\rm halo}}{\sum P_{mem}}))$ & $N_{halos}$ & $Log(\frac{M_{\rm halo}}{M_\odot})$ & $z$ \\
(1) & (2) & (3) & (4) & (5) & (6) & (7) & (8) \\
\hline
 $0.19\pm0.11$ & $-0.039\pm0.10$       & 0.11 & 0.14   & 0.14 &  179 & $14.01\pm0.24$  & 0.0-0.5\\
 $0.23\pm0.13$ & $-0.0069\pm0.14$       & 0.11 & 0.14   & 0.15 &  175 & $14.00\pm0.21$ & 0.5-0.75\\
 $0.25\pm0.14$ & $-0.0043\pm0.14$       & 0.12 & 0.17  & 0.18 &  245 & $13.90\pm0.21$  & 0.75-1.0\\
 $0.27\pm0.15$ & $0.0034\pm0.17$       & 0.12 & 0.19  & 0.21 &  229 & $13.84\pm0.22$ & 1.0-1.25\\
 $0.26\pm0.14$ & $-0.017\pm0.15$       & 0.12 & 0.17   & 0.18 &  159 & $13.79\pm0.19$ & 1.25-1.5\\
 $0.27\pm0.15$ & $0.0050\pm0.17$       & 0.15 & 0.19  & 0.21 &  127 & $13.73\pm0.20$  & 1.5-1.75\\
 $0.33\pm0.14$ & $0.061\pm0.17$       & 0.13 & 0.14  & 0.17 &  60  & $13.70\pm0.18$ & 1.75-2.0\\
 $0.29\pm0.13$ & $0.017\pm0.15$       & 0.13 & 0.16  & 0.18 &  33  & $13.68\pm0.19$ & 2.0-2.5\\
  \hline
\end{tabular}
\caption{Richness results for different redshift intervals. Column description. (1) Mean logarithmic difference between the estimated richness $\lambda_1=N_{\rm estimated}$ and the true richness $N_{\rm true}$, where the reported uncertainty is the rms dispersion around the mean and the richness $\lambda_1=N_{\rm estimated}$ refers to galaxies with $P_{mem}>P{thr}=20\%$; (2) as in Column~1, but here the estimated richness $\lambda_2=\sum P_{mem}$ is used; (3) rms dispersion around the mean of the logarithmic difference between the halo mass $M_{\rm halo}$ and the halo occupation number $N_{\rm true}$; (4) rms dispersion around the mean of the logarithmic difference between the halo mass $M_{\rm halo}$ and the estimated richness $\lambda_1=N_{\rm estimated}$; (5) as in Column~4, but here the estimated richness $\lambda_2=\sum P_{mem}$ is used; (6) number of halos considered; (7) median logarithmic halo mass, the reported uncertainty is the rms dispersion; (8) redshift bin.}\label{tab:richness_results}} 
\end{table*}

\begin{table*}[htbc]
\centering
{\small
\begin{tabular}{cccccccc}
\hline\hline
$\langle Log(\frac{N_{\rm estimated}}{N_{\rm true}})\rangle$  & $\langle Log(\frac{\sum P_{mem}}{N_{\rm true}}) \rangle$  & $\sigma(Log(\frac{M_{\rm halo}}{N_{\rm true}}))$ & $\sigma(Log(\frac{M_{\rm halo}}{N_{\rm estimated}}))$ & $\sigma(Log(\frac{M_{\rm halo}}{\sum P_{mem}}))$ & $N_{halos}$ & $Log(\frac{M_{\rm halo}}{M_\odot})$ & $z$ \\
(1) & (2) & (3) & (4) & (5) & (6) & (7) & (8) \\
\hline
 $0.24\pm0.13$ & $-0.030\pm0.16$       & 0.07 & 0.14   & 0.17 &  106 & 13.3-13.6 & $1.50\pm0.38$  \\
 $0.25\pm0.15$ & $-0.012\pm0.16$       & 0.09 & 0.17   & 0.18 &  565 & 13.6-13.9 & $1.15\pm0.45$  \\
 $0.26\pm0.14$ & $0.013\pm0.14$       & 0.09 & 0.16   & 0.17 &  412 & 13.9-14.2 & $0.83\pm0.42$  \\
 $0.21\pm0.11$ & $-0.012\pm0.11$       & 0.09 & 0.13   & 0.14 &  107 & 14.2-14.5 & $0.75\pm0.31$  \\
 $0.21\pm0.08$ & $-0.0086\pm0.068$       & 0.07 & 0.10   & 0.10 &  16 & 14.5-14.8 & $0.46\pm0.32$  \\
\hline
\end{tabular}
\caption{Richness results for different halo mass intervals. Column description. Columns 1-6 as in Table~\ref{tab:richness_results}; (7) logarithmic halo mass interval; (8) median redshift of the halos, the reported uncertainty is the rms dispersion.}\label{tab:richness_results2}} 
\end{table*}

\section{Summary and conclusions}\label{sec:conclusions}
\subsection{Method}
We have introduced a new method to perform statistically robust membership assignments to the galaxies 
in galaxy clusters using photometric redshift information over an unprecedented broad range of redshift ($z\sim0-2.5$) and halo mass ($Log(M_{halo})\simeq13-15$), thus extending 
previous studies \citep[e.g.,][]{brunner2000,george2011,rozo2009}. 

{ First { relative probabilities} are defined (see Section~\ref{sec:Pmem}). They correspond to the chance a galaxy has of occupying an optimal region in the parameter space that fiducial cluster members are associated with. The adopted parameters are the redshifts, the cluster centric distance, and the magnitude of the galaxy in a given reference band. Then {absolute probabilities} $P_{mem}$ are computed properly rescaling the {relative probabilities} assuming that in the limit of a very rich cluster $P_{mem}$ should tend to unity for fiducial cluster members with the highest {relative probabilities}. }

The method takes {the magnitude distribution of both cluster and field galaxies as well as the radial distribution of galaxies in clusters} into account using a non-parametric formalism (i.e., without assuming specific models) and relies on Bayesian inference to take photometric redshift uncertainties into account. 
{We successfully tested the method against $1,208$ galaxy clusters with at least ten members brighter than $\sim$0.25$L_\ast$ within $r_{200}$. The clusters have median redshift $z=1.00$ and median mass $10^{13.87}~M_\odot$. They span redshift and mass intervals $z=0.05-2.55$ and $10^{13.29-14.80}~M_\odot$, respectively. The clusters are drawn} from wide field simulated galaxy mock catalogs mainly developed for the forthcoming {\it Euclid} mission \citep{merson2013,Gonzalez-Perez2014}. Magnitude limited $\mathsf{H}<26$ galaxy mock catalogs { with different statistical photometric redshift accuracy $\sim(0.02-0.05)(1+z)$} are considered. The catalogs simulate the forthcoming (20.4~square degree) {\it Euclid} deep survey \citep{laureijs2011,laureijs2014}.

{ Membership probabilities are assigned to 22,906 galaxies, 9.8\% of them (2,236) are considered as {true} cluster members, which were identified when generating the simulations, according to the halo merging history \citep{merson2013}.}\\


\subsection{Testing membership probabilities}
Median purity and completeness values of ${(55^{+17}_{-15})}\%$ and ${(95^{+5}_{-10})}\%$ are reached for galaxies brighter than 0.25$L_\ast$ within $r_{200}$ of each simulated halo and for a statistical photometric redshift accuracy $\sigma((z_s-z_p)/(1+z_s))=0.03$. The mean values 
of purity (${\overline{\mathsf{p}}=56\%}$) and completeness (${\overline{\mathsf{c}}=93\%}$) are consistent with the corresponding median values and have negligible sub-percent uncertainties.

We observe stable mean values for both purity and completeness, with variations on the order of a few percent when the dependence on halo mass, richness, and cluster redshift accuracy are considered. 
{The largest departures from the mean values are found for} galaxies associated with distant $z\gtrsim1.5$ halos, faint {($\sim0.25\,L_\ast$)} galaxies, and those at the outskirts 
of the halo {(at cluster-centric projected distances $\sim r_{200}$)} for which the purity is lower,  {$\Delta \mathsf{p}\simeq20\%$ at most, with respect to the mean value}.

{{ We have tested the robustness of our membership probabilities $P_{mem}$ when used as absolute weights by means of the $P_{mem}$ vs. $f_{\rm true}$ scatter plot, where $f_{\rm true}$ is the fraction of true members among galaxies with membership probabilities $\sim P_{mem}$.  An overall agreement between the two quantities is found ($\chi^2/{\rm d.o.f.}=24.5/10=2.5$) at a few-percent accuracy, $\langle f_{\rm true} - P_{mem}\rangle = (-4.1\pm5.6)\%$.}}

{{ We suggest that our assignments can be used to estimate weighted cluster radial profiles and cluster galaxies luminosity functions, once the {decrease} of purity and completeness with radius is taken into account. }}\\


{{ We have verified that the few-percent  $P_{mem}$ vs. $f_{true}$ discrepancy is mainly due to the contamination of field galaxies at the outskirts {(at cluster-centric projected distances $\sim r_{200}$)}, where the $P_{mem}$ values are less reliable. Such results are consistent with the above mentioned significant decrease of the purity with increasing cluster-centric distance, also found in previous work \citep{george2011}.}}\\

{{ We have exploited the simulations to test the impact of correlated structures with masses $\gtrsim10^{13}~M_\odot$.
We found that when correlated structures are removed from both the cluster fields, 
{i.e., within the virial radius},
and the areas where the background is estimated the $P_{mem}$ bias disappears ($\langle f_{\rm true} - P_{mem}\rangle = (-0.87\pm9.1)\%$). However both the scatter (9.1$\%$) and associated $\chi^2/{\rm d.o.f.}=47.8/10$ are higher than the values obtained in the case where no correlated structure is removed. This is ultimately due to the fact that removing correlated structures implies higher-shot noise, which is critical in our case of low-number statistics and relatively poor clusters.}}

{ 
{We also suggest that our results could improve considering more massive clusters (work in preparation), where the contamination from field galaxies is less significant and/or including in the formalism the cluster radial profile (when known with sufficient accuracy) as additional prior information.
In particular, we checked that limiting ourselves to the subsample of 27 clusters with true richness $\geq40$ leads to a significant increase $\Delta\mathsf{p}\gtrsim10\%$ in purity.}
}\\

{ We have tested alternative strategies to define membership probabilities without rescaling. 
They refer to a different consideration of the redshift information. i) We enlarged the optimal region in the parameter space (concerning the redshift) when defining the {relative probabilities} using separately a top-hat and a Gaussian kernel, similarly to previous work \citep{brunner2000,rozo2015}. ii) We changed the redshift PDF of galaxies in the cluster field, 
{i.e., within the virial radius}, 
to account for the presence of the overdensity of photometric redshifts around the cluster redshift, similarly to previous work \citep{george2011}. In both cases i) and ii) probabilities on the order of unity are reached at maximum for fiducial cluster members, without the need of rescaling. 
However we verified 
that our formalism exploiting the rescaling leads statistically to better results than all tested cases.} \\

\subsection{Richness estimates}

The method is also applied to derive accurate richness estimates. A statistical comparison between the true ($N_{\rm true}$) vs. estimated ($\sum P_{mem}$) richness remarkably yields to unbiased results, $Log(\sum P_{mem}/N_{\rm true})=-0.0051\pm0.15$, where the reported values refer to all halos in the sample.
 
{ We have also exploited our richness estimates as halo mass proxy. The rms dispersion around the mean of the logarithmic difference between the halo mass and the estimated richness is $\sigma(Log(M_{\rm halo}/\sum P_{mem}))=0.20$. When massive halos $\geq10^{14.5}~M_\odot$ are considered the scatter is reduced to 0.10~dex. Our results are fairly consistent with $\sigma(Log(M_{\rm halo}/\sum P_{mem}))\simeq\sqrt{2}\sigma(Log(M_{\rm halo}/N_{\rm true}))$, where $\sigma(Log(M_{\rm halo}/N_{\rm true}))$ refers to the (purely cosmological) scatter of the halo occupation distribution. }\\


Our estimates could be useful to constrain the cluster mass function as well as calibrate independent cluster mass estimates such as those obtained from weak lensing, Sunyaev-Zel’dovich, and X-ray studies. 
Membership assignments derived with our method could be used to perform studies of evolution of galaxies in clusters. Our method can be applied to any list of galaxy clusters or groups in both present and forthcoming surveys such as SDSS, CFHTLS, Pan-STARRS, DES, LSST, and {\it Euclid}.

 \begin{acknowledgements}
{ We thank the anonymous referee for very helpful comments.}
{ We thank Violeta Gonzalez-Perez, Alex Merson, Carlton Baugh, and Peder Norberg for kindly providing their simulations products which this work is based on and which was created for the {\it Euclid} Consortium. We thank them for helpful discussion about their correct use and properties. These simulations products were created on the DiRAC Data Centric system at Durham University, operated by the Institute 
for Computational Cosmology on behalf of the STFC DiRAC HPC Facility (www.dirac.ac.uk). This equipment 
was funded by BIS National E-infrastructure capital grant ST/K00042X/1, STFC capital grant ST/H008519/1 
and ST/K00087X/1, and STFC DiRAC Operations grant ST/K003267/1 and Durham University. 
DiRAC is part of the National E-Infrastructure.}
We thank Chiara Ferrari, Alberto Cappi, R\'{e}mi Adam, and Sophie Maurogordato for helpful scientific discussion.
We thank Pier-Francesco Rocci, Srivatsan Shridar, and Martin Vannier as well as Guillame Guillon, Alvaro Alonso Rojas Arriagada, and Marina Ricci for helpful discussion about shell scripting and Python coding.
\end{acknowledgements}

\end{document}